\documentclass[conference]{IEEEtran}
\usepackage{cite}
\usepackage{amsmath,amssymb,amsfonts}
\usepackage{algorithmic}
\usepackage{graphicx}
\usepackage{textcomp}
\usepackage{xcolor}
\usepackage{fancyhdr}
\usepackage[hyphens]{url}

\def\BibTeX{{\rm B\kern-.05em{\sc i\kern-.025em b}\kern-.08em
    T\kern-.1667em\lower.7ex\hbox{E}\kern-.125emX}}

\fancypagestyle{firstpage}{
  \fancyhf{}

  \fancyfoot[C]{\thepage}
}

\title{Direct Spatial Implementation of Sparse Matrix Multipliers for Reservoir Computing} 
\author{\IEEEauthorblockN{Matthew Denton and Herman Schmit}
 \IEEEauthorblockA{Google Brain\\\{myuzaki, schmit\}@google.com}}

\usepackage[ruled,vlined]{algorithm2e}
\usepackage{listings}
\usepackage{xcolor}

\definecolor{codegreen}{rgb}{0,0.6,0}
\definecolor{codegray}{rgb}{0.5,0.5,0.5}
\definecolor{codepurple}{rgb}{0.58,0,0.82}
\definecolor{backcolour}{rgb}{0.95,0.95,0.92}

\lstdefinestyle{mystyle}{
    backgroundcolor=\color{backcolour},   
    commentstyle=\color{codegreen},
    keywordstyle=\color{magenta},
    numberstyle=\tiny\color{codegray},
    stringstyle=\color{codepurple},
    basicstyle=\ttfamily\footnotesize,
    breakatwhitespace=false,         
    breaklines=true,                 
    captionpos=b,                    
    keepspaces=true,                 
    numbers=left,                    
    numbersep=5pt,                  
    showspaces=false,                
    showstringspaces=false,
    showtabs=false,                  
    tabsize=2
}

\begin{document}
\maketitle
\thispagestyle{firstpage}
\pagestyle{plain}

\begin{abstract}
Reservoir computing systems rely on the recurrent multiplication of a very large, sparse, fixed matrix. We argue that direct spatial implementation of these fixed matrices minimizes the work performed in the computation, and allows for significant reduction in latency and power through constant propagation and logic minimization. Bit-serial arithmetic enables massive static matrices to be implemented. We present the structure of our bit-serial matrix multiplier, and evaluate using canonical signed digit representation to further reduce logic utilization. We have implemented these matrices on a large FPGA and provide a cost model that is simple and extensible. These FPGA implementations, on average, reduce latency by 50x up to 86x versus GPU libraries. Comparing against a recent sparse DNN accelerator, we measure a 4.1x to 47x reduction in latency depending on matrix dimension and sparsity. Throughput of the FPGA solution is also competitive for a wide range of matrix dimensions and batch sizes. Finally, we discuss ways these techniques could be deployed in ASICs, making them applicable for dynamic sparse matrix computations.
\end{abstract}

\section{Introduction}
The progress of machine learning algorithms and accelerators are bidirectionally linked. Algorithms that are well-structured for available computing achieve higher quality at less cost. \cite{tan2020efficientnet} Similarly, accelerators are benchmarked on pre-existing ML algorithms and succeed or fail accordingly.\cite{reddi2020mlperf} In this paper, we try to escape this self-reinforcing system by considering a different form of machine learning, contemplating what kind of accelerator this approach would require, and consider whether the results have consequences for conventional machine learning systems.

Current ML accelerators use matrix multiplication as the basic building block. These matrix multiplication units are primarily:
\begin{itemize}
\item Dense: Dense matrix multiplication is very efficiently implemented in silicon with systolic arrays. \cite{tpu} Algorithmically, density is a result of backpropagation, which tends to fill in the value of every weight with a number, even if that value means little to the end computation.  
\item Small: Dense matrix multipliers are easily composable to provide the computation for arbitrarily sized matrices. Therefore, a smaller matrix multiplier can be used as an efficient atom in a big matrix computation, whereas a big matrix multiplier will suffer from lower utilization on small matrix computations.
\item Two-operand: Finally, most matrix multipliers have both the matrix and the vector as stored variables. There is a need to change matrix values during training and in order to enable time-multiplexing of the multiplier unit to support different matrix dimensions. Having two operands can be costly from area and power. The primary power-saving technique for these multipliers is batching, where the matrix values are kept constant while performing multiple vector times matrix computations.
\end{itemize}

Reservoir computing, specifically Echo State Networks as discussed in this paper, uses very large, sparse matrices. These matrices are generated randomly and are never modified by training, and therefore the matrix is fixed for the lifetime of the computation. In order to handle these matrices, conventional ML accelerators perform indexing and tiling of the sparse matrix, which effectively transform the large sparse operation into multiple small dense operations. However, this transformation has compute and storage overhead, and tiling increases the latency of the computation. \textbf{This paper shows that it is both possible and advantageous to implement these large sparse operations directly in programmable hardware, which eliminates the cost of indexing and tiling.} We use Reservoir Computing as a motivating factor, but discuss how this could impact ML inference acceleration in the future.

\subsection{Contributions}

The primary contributions of our paper are as follows:
\begin{itemize}
\item We present a bit-serial architecture for fixed vector-matrix multiplication, which minimizes hardware, energy and latency.
\item We implement very large, sparse matrices using this architecture without paying the cost of indexing or tiling.
\item We present simple cost and power models, which enable the quick estimation of size and power of any fixed matrix on an FPGA.
\item We show how a simple transformation can significantly improve the achievable density.
\item We compare the achievable performance against state of the art sparse libraries for GPUs and DNN accelerators.
\item We discuss how this idea could be applied to a custom circuit that would eliminate the limitations of the FPGA.
\end{itemize}

\subsection{Outline}
The remaining content is structured as follows: 
In Section \ref{sec:rescomp}, we describe reservoir computing in detail and enumerate prior art. Section \ref{sec:bitserial} reviews the state-of-the-art in bit-serial machine learning acceleration and describes our low-latency bit-serial matrix multiplier. Section \ref{sec:synthesis} describes synthesis of our architecture and presents a simple area model based on our data. Section \ref{sec:csd} investigates the use of canonical signed digits in our matrices. Section \ref{sec:physical} discusses experiments compiling very large matrices onto a large FPGA and measuring area, frequency and power. Section \ref{sec:eval} compares these results against the state-of-the-art sparse libraries for comparable GPU implementations. We also compare against a sparse DNN accelerator. Finally, in Section \ref{sec:future} we discuss the implications of this approach to implementations of more conventional models and ASICs. 

\section{Reservoir Computing}
\label{sec:rescomp}
Reservoir computing is a subset of supervised machine learning which seeks to learn a non-linear transformation from an input sequence to an output sequence. It's conceptually similar to a recurrent neural network, but has one key difference, the recurrent weights are not trained. A reservoir system can be described using the following equations, adapted from \cite{TANAKA2019100}:
\begin{equation}
  \mathbf{x}(n) = f(W^{in}\mathbf{u}(n) + W\mathbf{x}(n-1))
\end{equation}
\begin{equation}
  \mathbf{y}(n) = W^{out}\mathbf{x}(n)
\end{equation}
Where \begin{math}\mathbf{u}(n),\mathbf{y}(n)\end {math} are the input and output at time $n$ respectively, \begin{math} W^{in}, W^{out} \end{math} are the input and output weight matrices, and \begin{math} f() \end{math} is a non-linear activation function. \begin{math}W\end{math} is a very large, fixed matrix, and $x$ is a large vector that carries the state information from one time unit to another.

Ordinarily, \begin{math} W, W^{in}\end{math} are initialized according to some heuristics, and remain fixed.  \begin{math} W^{out} \end{math} is trained via linear regression. It is evident to see why this approach is appealing from a compute perspective - only a linear regressor needs to be trained which completely eliminates the need for error backpropagation and allows the designer to choose the optimizer for the linear layer, which may be gradient-descent, least-squares, or any other optimization technique. 

A comparison of various systems for recognition of multivariate time series is presented in \cite{bianchi20}. This paper compares reservoir computing systems against fully trainable recurrent networks. The reservoir solutions train much faster with comparable quality. In this paper, the baseline reservoir computing system consists of a fixed square matrix with a dimension of 800 with 75\% of the elements being 0, which we henceforth refer to as "element sparsity".

A comprehensive summary of physical systems used to implement physical reservoirs \cite{TANAKA2019100} includes a section on FPGA implementations. The earliest citation is \cite{stochastic}, which describes encoding neuron values as "stochastic bitstreams". \cite{res_pca} describes an FPGA implementation of RC and presents heuristics for defining the input matrix \begin{math} W^{in}\end{math}. \cite{cyclic-res} describes an encoding system that allows for multiplication to be represented as shifts.

An FPGA design that includes both the reservoir as well as the output layer with auto-regression was presented in \cite{antonik2015fpga}. This system was used to perform channel equalization, which is ideal for online learning because the known patterns with expected results are presented on a periodic basis. This paper is not easily replicated, but the results state that the design used most of the block RAM on an FPGA, with low logic utilization. This paper did not emphasize performance exploration, as evidenced by the 33MHz clock frequency.

It is shown in \cite{kleyko2020integer} that the elements of these reservoirs can be quantized into integers. The authors show a variety of tasks where a precision of 3-4 bits leads to no accuracy loss. They evaluate their design on a ZedBoard FPGA with an integrated ARM CPU and are able to run a 300x300 reservoir at 100MHz. Studies in \cite{gallicchio2020sparsity} show that sparsity should exceed 80\% to maximize performance and enable rich interaction among neurons.

The idea of using large, sparse matrices has consequences beyond reservoir computing. Numenta~\cite{numenta2019} is a company creating  artificial neural networks by mimicking biological networks, which are all sparse. They are also using large, sparse, random, fixed matrices with a trainable output selection layer for their investigations. Finally, \cite{lth18} showed that special subnetworks with 80-95\% sparsity exist within backpropagation-trained dense networks. These subnetworks have the same or better error rates than full dense networks. In other words, most of the computation performed in inference using the full matrix is wasted.

The remainder of this paper details our approaches to implement a large, fixed matrix multiplication on an FPGA using bit serial arithmetic. Specifically, we accelerate the key primitive in reservoir computing:
\begin{equation}
\label{eqn:prim}
    o = a^{\intercal}V 
\end{equation}
Which is often called "gemv" and is an important primitive for many machine learning workloads.

\section{Bit-serial vector-matrix multipliers}
\label{sec:bitserial}
Bit-serial arithmetic was an important technique decades ago when parallel arithmetic was too costly.  Simple VLSI implementations of bit-serial neural networks \cite{bitserialnn} were proposed as far back as 1988.  More recently, DNN accelerators have used bit-serial arithmetic to support variable precision arithmetic.
STRIPES \cite{stripes} presents a bit-serial architecture that enables different precisions on a per-layer basis, quantizing each layer to their required bit-widths.  Bit-fusion \cite{bit-fusion} enables dynamic-precision for inputs as well.  Bit-pragmatic \cite{bit-pragmatic} and Laconic \cite{laconic} exploit bit-level sparsity. They detect the bits that are zero and skip those multiplications. In all the above cases, the DNN dimensions often exceed the compute capability of the hardware, so tiling and striding the inner product is regularly employed.

As discussed earlier, reservoir computing uses a single large, sparse, weight matrix that is fixed for both training and inference. This allows us to optimize the logic implementation by performing constant propagation on the matrix to remove the logic corresponding to all zeros; both zeros as whole terms, and bits that are zeros within individual terms. It would be possible to implement these matrices in custom logic (ASIC), but in this work we use an FPGA. The FPGA consists of an array of Lookup Tables (LUTs), Registers (FFs), and programmable interconnect. FPGAs generally include other programmable resources such as embedded multipliers and SRAMs, which we will not consider in this work. The particular FPGA we are using has the capability to re-purpose some of the LUTs into small RAMs or shift registers which are called LUTRAMs.

Our FPGA design flow takes the content of the matrices and compiles it to a physical design consisting of the programmable logic resources and interconnect. This design flow produces an achievable frequency, area, and power estimation.

\subsection{The Microarchitecture}
\begin{figure}[h]
  \centering
  \includegraphics[width=\linewidth]{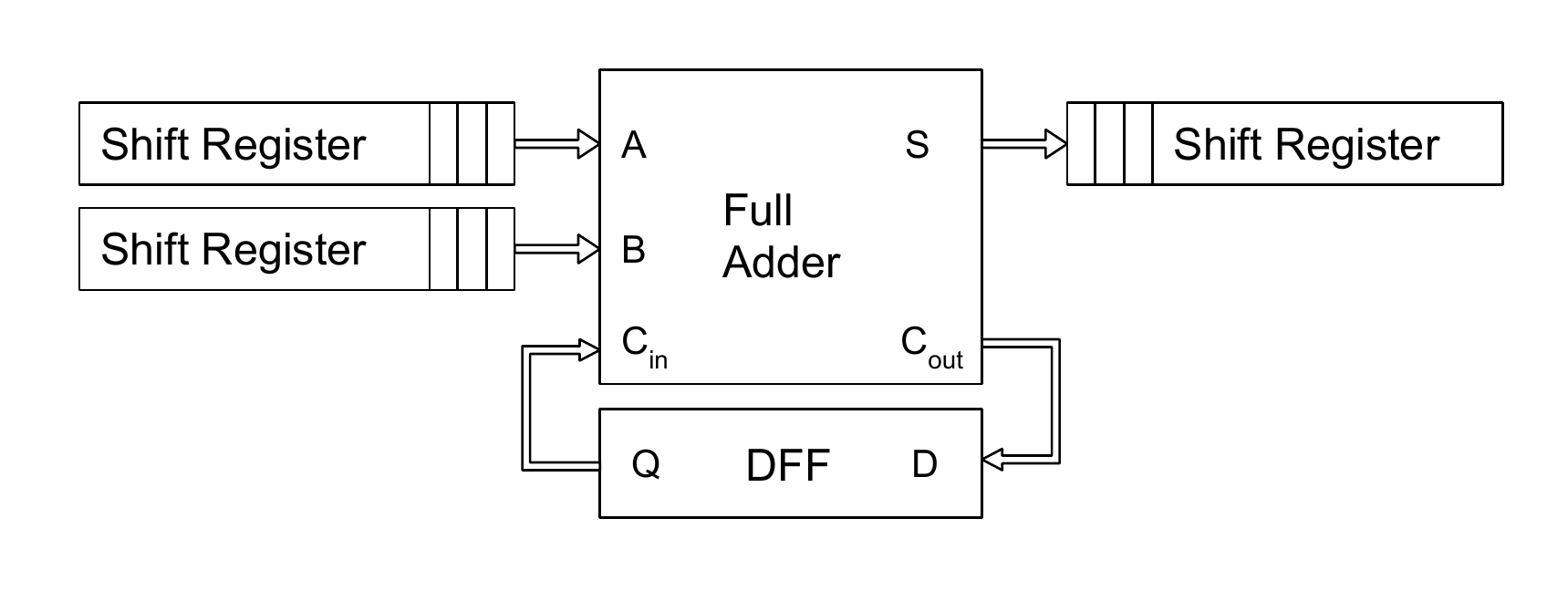}
  \caption{Bit-serial adder}
  \label{fig:bs-adder}
\end{figure}

Figure \ref{fig:bs-adder} shows the design of a bit-serial adder, which is the basic building block of bit-serial arithmetic units. 

In the bit-serial adder, inputs A and B are shifted into a full-adder one bit each clock cycle, LSb first. The result of the addition (S) is shifted into a result register. The carry is registered as the carry input for the next cycle. This is analogous to manual long addition - add one digit at a time, with any resulting carry being added in the next most significant digit. Rather than chaining the carries through multiple adders "in space", we use a single adder and process the carry "in time". Table \ref{tab:bs_exl} shows an example of a bit-serial addition of \begin{math} 3_{10} + 7_{10} = 10_{10} \leftrightarrow 011_{2} + 111_{2} = 1010_{2} \end{math}

\begin{table}[h]
  \centering
  \caption{Bit-serial Addition Example}
  \label{tab:bs_exl}
  \begin{tabular}{c|cccccc}
    Cycle & $C_{in}$ & A & B & S & $C_{out}$ & Result\\
    \hline
    1 & 0 & 1 & 1 & 0 & 1 & 0000\\
    2 & 1 & 1 & 1 & 1 & 1 & 1000\\
    3 & 1 & 0 & 1 & 0 & 1 & 0100\\
    4 & 1 & 0 & 0 & 1 & 0 & 1010\\
\end{tabular}
\end{table}

Similarly, we can create a bit-serial subtractor that performs $a - b$ by initializing the carry bit to 1, and adding a NOT gate between $b$'s register and the full adder. This scheme is equivalent to negating $b$ in two's complement and then adding it to $a$. In the FPGA, the bit serial adder or subtractor can be mapped to a single 6-input LUT and two registers. The timing path that determines the frequency for this design (the critical path) will be very short: a single stage of logic, the setup to the flop, and any interconnect delay between the components. 
\paragraph{Single-bit dot-product} Using these primitives, we now build a single-bit dot-product, as described below:

\begin{equation} 
\label{eqn:1-bit_prot_prod}
s =\sum_{i=1}^{N} a_{i}*b_{i}
\end{equation}

Where $a, b$ are N-length vectors whose elements are 1-bit wide, and $s$ is a scalar integer. Because we are multiplying single bits, we can realize the multiplication with a simple AND gate. After multiplying each pair of elements, we reduce the partial sums through a tree of bit-serial adders.

\begin{figure}[h]
  \centering
  \includegraphics[width=\linewidth]{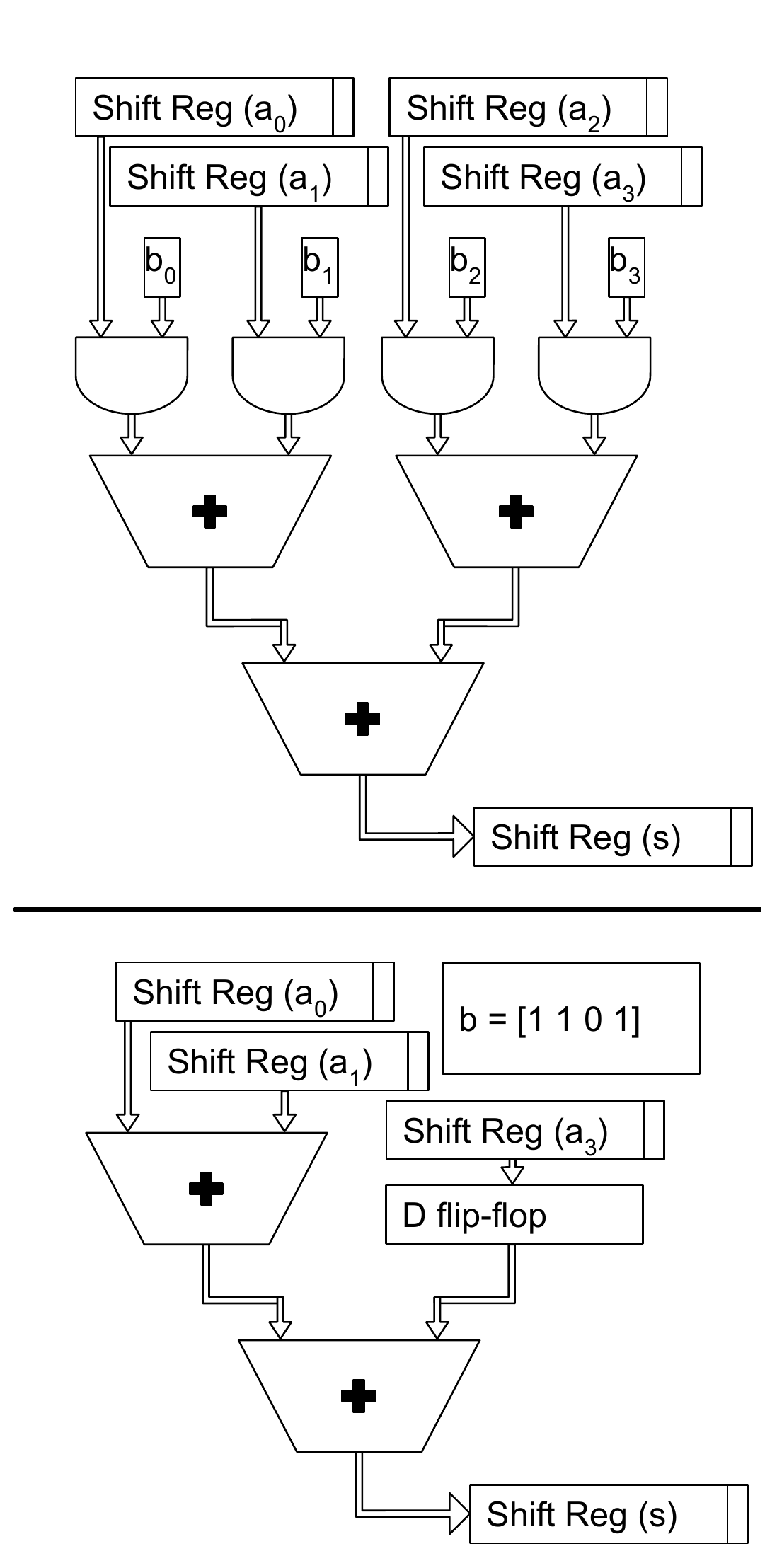}
  \caption{a)Multi-bit by fixed single-bit dot-product with tree reduction and shift registers (N=4) (trapezoids represent bit-serial adders, which have flip-flips between each stage but are omitted for brevity)\\ b) Setting b = [1 1 0 1] reduces the circuit}
  \label{fig:mixed-bit_dot_prod}
\end{figure}

\paragraph{Multi-bit input} We can extend this design to support one of the operands, $a$, being multiple bits wide. We hold the single-bit vector $b$ fixed, and stream the multi-bit vector through the circuit one bit at a time from LSb to MSb using a shift register for each dimension. Similarly, we capture the result in a shift register. One can think of this circuit as summing the elements of $a$ that are selected by a Boolean vector $b$. This design is shown in Figure \ref{fig:mixed-bit_dot_prod}a.

Note that because this is merely summing the dimensions selected by $b$, this circuit works for both signed and unsigned $a$ inputs. To ensure signed inputs produce the correct sign bit, we sign extend the input $a$ from the shift register until the computation has finished. 

Given that our boolean vector $b$ is fixed, consider a given element $b_i$. If this element is set to 1, the input to the AND gate is 1, so the output of the AND gate will just be the associated $a_i$. In this case, we can cull the AND gate entirely and connect the input $a_i$ directly to the bit serial adder. On the other hand, if the element is 0, the AND gate will have a fixed 0 input and therefore its output will always be 0. Furthermore, one of the inputs to the bit-serial adder is now a fixed 0, so the adder is just passing the value forward. In this case, the adder is acting as a D-flip-flop. This means we can not only cull the AND gate, but we can also replace the adder with a single flip-flop, which reduces the cost significantly. Put simply, we can greatly reduce the cost of this circuit by fixing $b$, and the cost should be proportional to the number of bits set in $b$. \textbf{This optimization is the fundamental minimization technique in our design.} An example of this reduction can be seen in Figure \ref{fig:mixed-bit_dot_prod}b. We will revisit this idea in Section \ref{sec:synthesis}.

\paragraph{Multi-bit input and weight} Finally, we can combine multiple of these multi-bit by 1-bit dot-products to create a full multi-bit dot-product. This allows the dot-product of two integer vectors, each with its own bit-width. Because each dot-product in the previous circuit occurs at the bit-level in $b$, we merely need to create such a circuit for each bit position in $b$, and then combine the results for each bit position.

\begin{figure}[h]
  \centering
  \includegraphics[width=\linewidth]{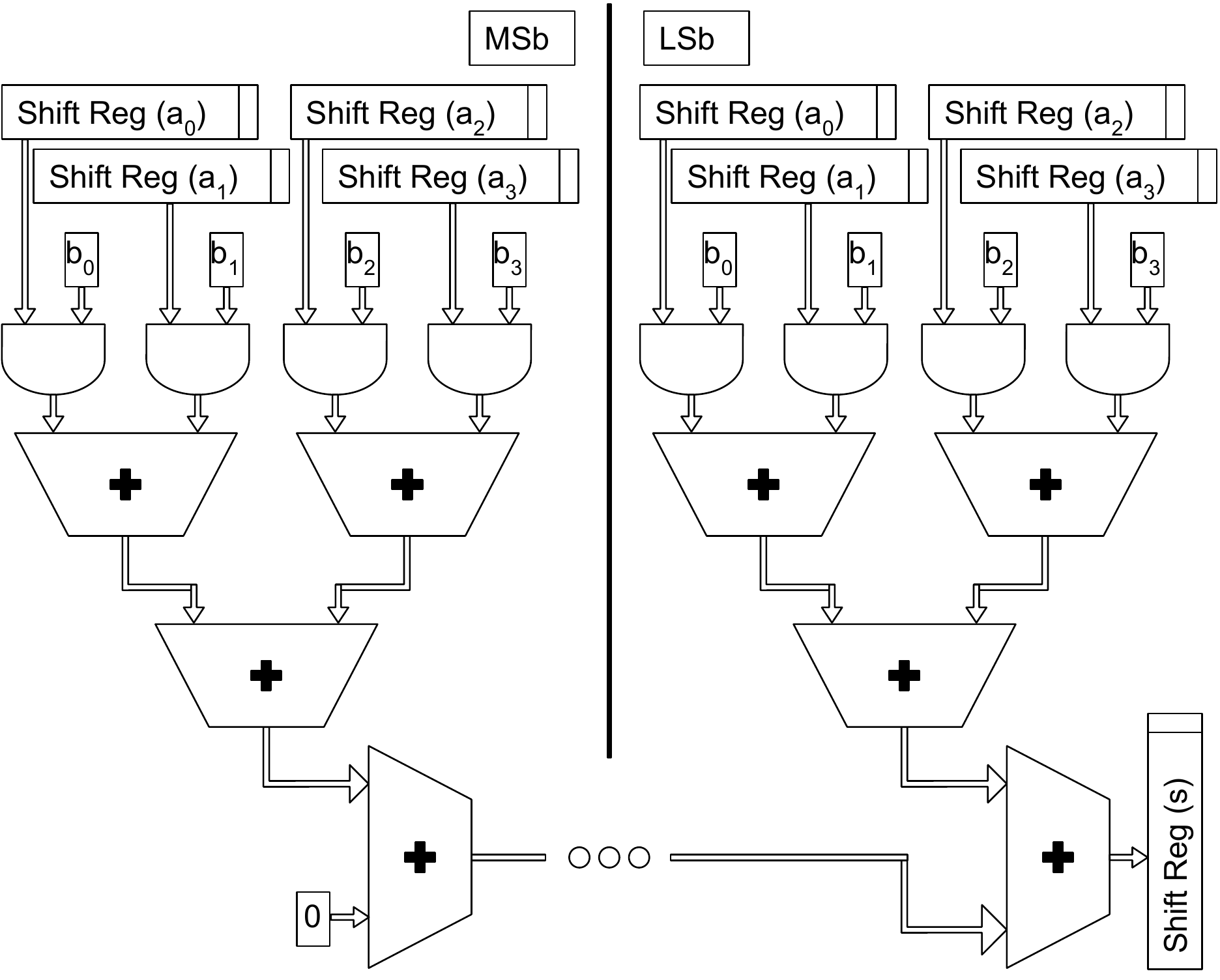}
  \caption{Multi-bit Dot-product }
  \label{fig:multi-bit}
\end{figure}

Luckily, the bit-serial adders already have this capability by shifting in time. For example, to make a 2-bit dot-product circuit, we make one circuit for the MSb, and one for the LSb. The results of the two are combined by feeding the LSb result into a bit-serial adder, and then feeding the MSb result, delayed by 1 cycle, into the same bit-serial adder, which then outputs into a shift register to capture the final result. Delaying the MSb result by one cycle effectively multiplies it by 2. To extend this idea to arbitrary bit-width, we just create a chain of bit-serial adders, where the bottom input is the previous link in the chain, and the top input is the corresponding bit. In this case, the result of a bit position is delayed accordingly. A multi-bit example is shown in Figure \ref{fig:multi-bit}. Notice that the MSb is fed into a bit-serial adder along with 0, which becomes a D flip-flop.

This architecture works for signed inputs and unsigned inputs, but the weights are unsigned. An easy way to implement signed weights is to separate the positive and negative terms of the $b$ vector into two separate unsigned vectors, and simply subtract the two resultant streams. Because the number of ones in the two matrices is conserved by this transform, it makes almost no impact on the total area, and adds a single cycle to the latency. 

\begin{figure}[h]
  \centering
  \includegraphics[width=\linewidth]{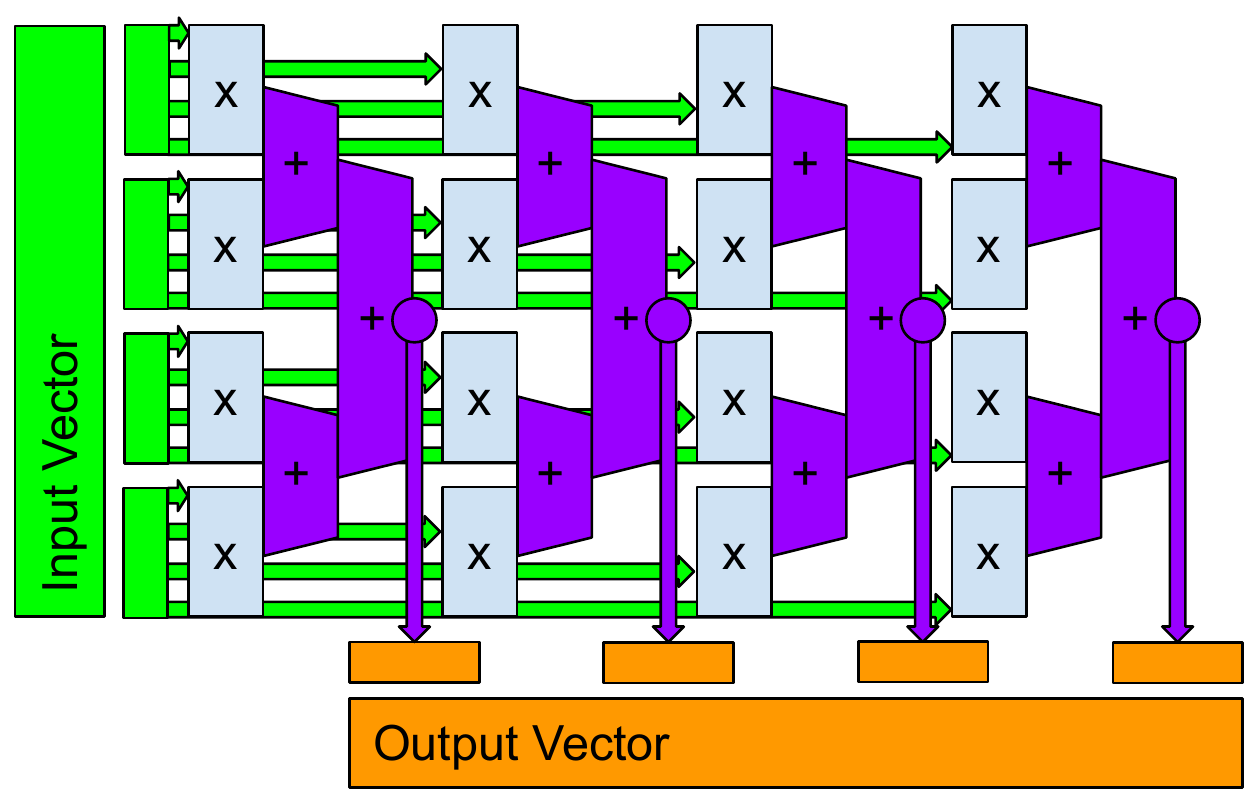}
  \caption{Full Vector-Matrix Multiplier Architecture \\ Broadcast of inputs shown in green. Multiplication at bit-level of fixed weights shown in blue. Per-column partial sum reduction shown in purple, yielding the final output vector in orange}
  \label{fig:arch}
\end{figure}

\paragraph{Vector-matrix product} Now that we have a unit to perform dot-products, we can finish up this matrix-multiplier by creating a dot-product unit for each column in the matrix. Then, we just broadcast the input on each row to all columns, and that results in a matrix multiplier, which takes shape as shown in Figure \ref{fig:arch}. (Full bit-serial circuits are omitted here for brevity)

Because the computations are parallelized across the matrix columns, and then again across each bit-position, the latency for this design is described in Equation \ref{eqn:latency}:

\begin{equation}
    \label{eqn:latency}
    \text{Latency}=BW_{i} + BW_{w} + \log_2 R + 2
\end{equation}
Where $BW_{i}$ and $BW_w$ are the bitwidth of the input and weights respectively, and $R$ is the number of rows in the matrix. We incur the input width to stream the input in, the output width to stream the output out, and our adder tree is logarithmic in depth. We incur a single cycle to accumulate across bit positions and an additional cycle to subtract the positive and negative weight matrices. For example, given 8-bit inputs and weights and a 1024x1024 weight matrix, we perform the vector-matrix product in $8 + 8 + \log_2(1024) + 2 = 28$ cycles.  

\section{RTL Synthesis Results}
\label{sec:synthesis}

In this section, we seek to understand the behaviour of our design by studying it on small matrices. We extend the insights here to larger matrices in Section \ref{sec:physical}. Recall the optimization from Section \ref{sec:bitserial} that if a bit in the weight-matrix is 0, we can cull the AND gate and bit-serial adder, replacing it with a simple D flip-flop.  We coded our design in SystemVerilog and ran synthesis in Xilinx Vivado 2020.2 targeting Xilinx UltraScale+ devices \cite{ultrascale-plus}. If our design behaves as expected, the hardware cost should be proportional to the number of bits set in the weight matrix. To test this, we randomly initialized a series of 64x64 weight matrices at 8-bit precision. For each bit in the weight matrix, we sample from a Bernoulli distribution, where the $p$ parameter is equal to (1 - bit\_sparsity). That is to say, the bit-sparsity of the weight matrix is the number of bits that are 0 out of the total number of bits. 0\% bit-sparse means all bits are 1, 50\% means the bits are uniformly random between 0 and 1, and 100\% means all bits are 0. This gives us a straightforward way to measure the cost of these matrices.

\begin{figure}[h]
  \centering
  \includegraphics[width=\linewidth, scale=0.7]{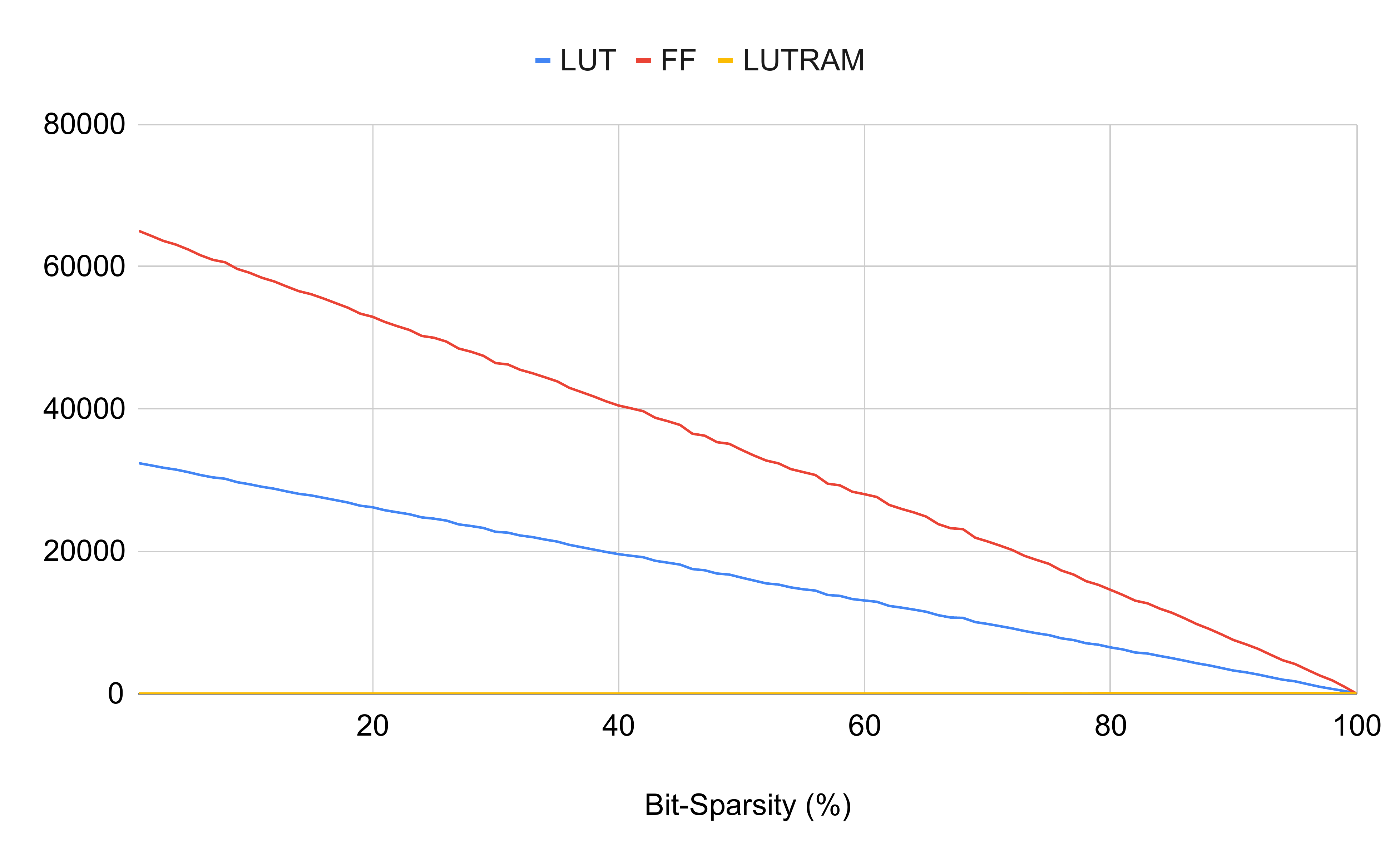}
  \caption{Hardware utilization vs bit-sparsity of a 64x64 matrix}
  \label{fig:utilvsparsity}
\end{figure}

Figure \ref{fig:utilvsparsity} shows synthesis results for sweeping bit-sparsity from 0\% to 100\%. We report utilization of LUTs, FFs, and LUTRAMs. 

As expected, the total hardware cost of our architecture is linear with respect to the number of bits set in the weight matrix. 

\begin{figure}[h]
  \centering
  \includegraphics[width=\linewidth]{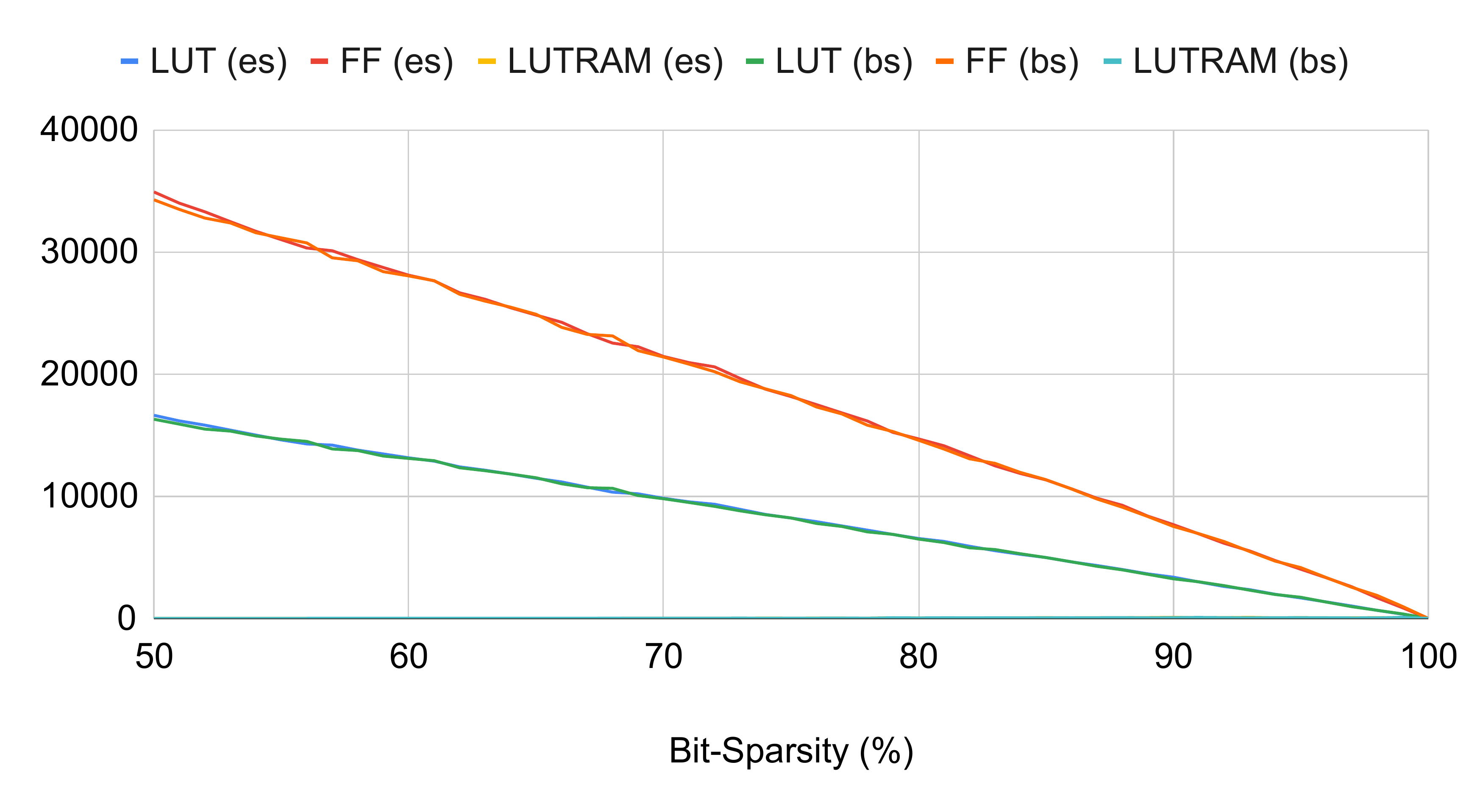}
  \caption{Cost of Element Sparse Matrix Compared To Bit-Sparse Matrix \\ 
  Here, (es) and (bs) designate Element Sparse and Bit-Sparse schemes respectively. While results do not match exactly, they are within an acceptable noise margin}
  \label{fig:EsBs}
\end{figure}

This illustrates that our architecture effectively exploits bit-level sparsity to reduce the cost of fixed-matrix multiplication. Recent advances in neural-networks and architecture have similarly sought to exploit element-level sparsity to make computations more efficient. \cite{han2016deep, han2016eie} By element-level sparsity, we mean a portion of the weights being 0, so there is no need to multiply by them. Because bit-level sparsity is a super-set of element-level sparsity, we wanted to know if our design was ineffective given an element-sparse matrix. To test this, we randomly initialized a set of matrices, where the weights are sampled from a uniform distribution of all possible values for the given bit-width. In this case, the matrix is 50\% bit-sparse, as every bit has an equal probability of being 0 or 1. We then randomly replace matrix elements with 0 until we reach a desired level of element-sparsity. Taking these samples, we convert the element-sparse value into a bit-sparse value, and compare the two approaches. This second experiment encourages bits to gather in individual elements, rather than the bit-sparse experiment which encouraged bits to be spread out. Figure \ref{fig:EsBs} shows our findings and plots them against the original experiment.  

The graph shows us that it doesn't matter if the bits are concentrated or not. The two lines are nearly identical, which means that we don't have to make any concessions to support element-sparse designs. Unlike most sparse accelerators, this design exploits sparsity in the elements when matrix entries are zero, and also elicits great benefit from elements being powers-of-two. 

\begin{figure}[h]
  \centering
  \includegraphics[width=\linewidth, scale=0.7]{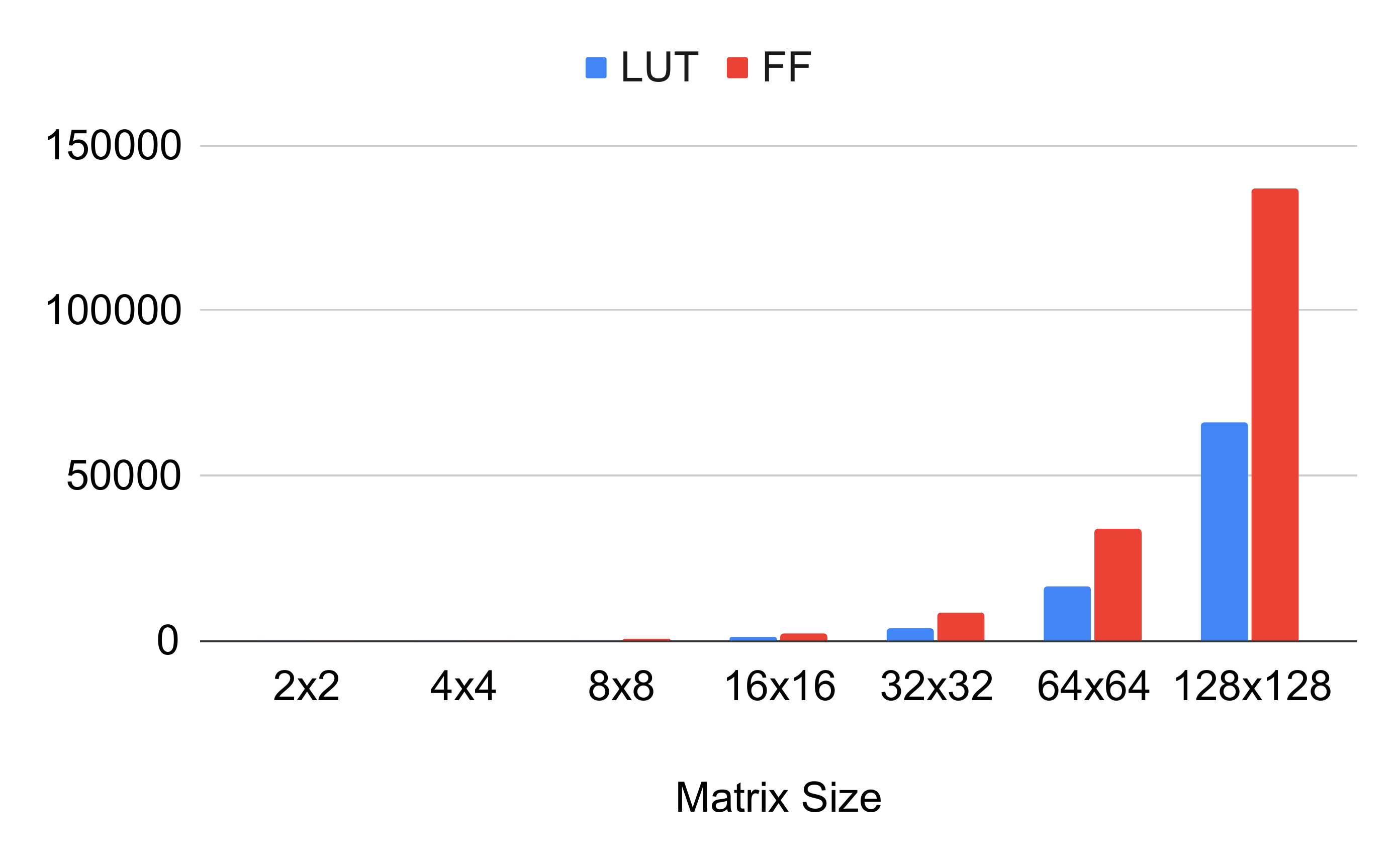}
  \caption{Hardware Utilization vs Matrix Size for random 8-bit integers}
  \label{fig:mat_size}
\end{figure}

Figure \ref{fig:mat_size} shows the LUT and FF utilization vs matrix size. The cost is quadratic with respect to matrix dimension and therefore linear with respect to the number of elements. This implies there is little optimization our RTL flow is doing across weight elements. In other words, large matrices are no more and no less dense than smaller matrices. 

\begin{figure}[h]
  \centering
  \includegraphics[width=\linewidth]{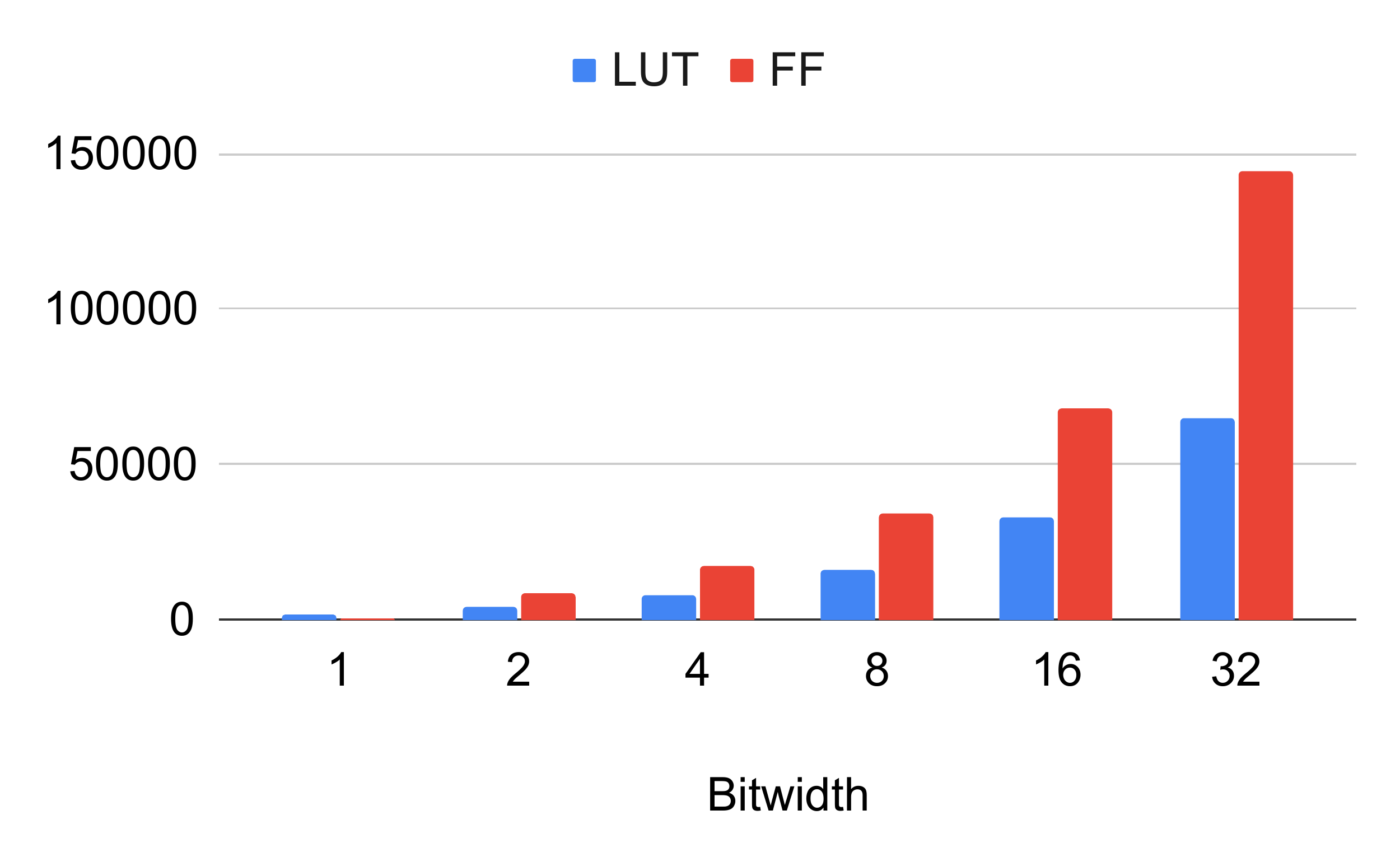}
  \caption{Hardware Utilization of 64x64 Random Matrix for Varying Bitwidths}
  \label{fig:bitwidth}
\end{figure}

We also explore the cost function with respect to the bit-width of the weights. Figure \ref{fig:bitwidth} shows the cost of a 64x64 matrix, sweeping the weights bit-width from 1-bit to 32-bit. Because the architecture performs 1-bit dot-products for each bit-position, and then accumulates the results, we would expect there to be little cross-bit optimizations. Indeed, we observe a linear LUT and FF cost with respect to the bit-width of the weights.

\section{Canonical Signed Digit}
\label{sec:csd}
In an effort to reduce the cost of the multiplier, we consider Canonical Signed Digit (CSD) representations \cite{signed_digit}. CSD decomposes an unsigned integer into a sum of a positive and negative integers, where the positive and negative integer together have fewer set bits than the original number. For example, consider the following :
\begin{equation*}
    15_{10} = 16_{10} - 1_{10} \leftrightarrow 1111_{2} = 10000_{2} - 00001_{2}
\end{equation*}
Notice that we can decompose a number that had four bits set to the difference of two numbers which each have one bit set, for a total of two. Do also take note that the bit-width of the decomposition is one wider than the original. Recall that the cost function of our multiplier is the number of bits set, so this shows promise in our case. We transform Equation \ref{eqn:prim} using:
\begin{equation}
\label{eqn:csd}
    V \equiv P - N \implies o = a^{\intercal}(P-N) = (a^{\intercal}P) - (a^{\intercal}N) 
\end{equation}
To transform V into N and P, we apply the algorithm in Listing \ref{lst:csd} element-wise, placing positive elements in P, and placing the absolute value of negative elements in N. In short, the algorithm searches for strings of consecutive 1 bits, which we call a chain. If a chain is length 1, nothing is done. If a chain is length 2, we flip a coin. On heads, we replace the chain with a +1 in the bit position one-past that of the MSb in the chain and a -1 at the LSb in the chain. On tails, we leave the original representation. For chains of length 3 and greater, we perform the same replacement that we do for heads on length 2. We introduce the random variable to balance the decomposition, since a transformation of a length 2 chain has no benefit and no detriment. 

When operating on signed weights, we perform a CSD transform on both the positive and negative weight matrices. Positive elements that result from CSD remain in the original matrix, and negative elements are transferred to the opposite weight matrix.  

\begin{lstlisting}[language=Python, caption= CSD Conversion Algorithm, label=lst:csd]
def convert_to_csd(num_bin_list):
  local_list = list(num_bin_list)
  target = [0] * (len(local_list) + 1)
  local_list.reverse()  
  chain_start = -1  # are we in a chain?
  for i in range(len(target)):
    if i < len(local_list):
      bit = local_list[i] 
    else: 
      bit = 0 
    if bit == 0:
      if chain_start == -1:  # no chain
        target[i] = 0  # nothing to be done here
      else:
        #  We terminate a chain, how long is it?
        chain_length = i - chain_start
        if chain_length == 1:  # leave it alone
          target[chain_start] = 1
        elif chain_length == 2:  # a chain of two
          if bool(random.getrandbits(1)):
            # do the substitution
            target[chain_start] = -1
            target[i] = 1
          else:
            target[chain_start] = 1
            target[i-1] = 1
        else:  # will get benefit
          target[chain_start] = -1
          target[i] = 1
      chain_start = -1  # not in a chain anymore
    else:  # bit == 1
      if chain_start == -1:
        chain_start = i
  target.reverse()
  return target
\end{lstlisting}

\begin{figure}[h]
  \centering
  \includegraphics[width=\linewidth]{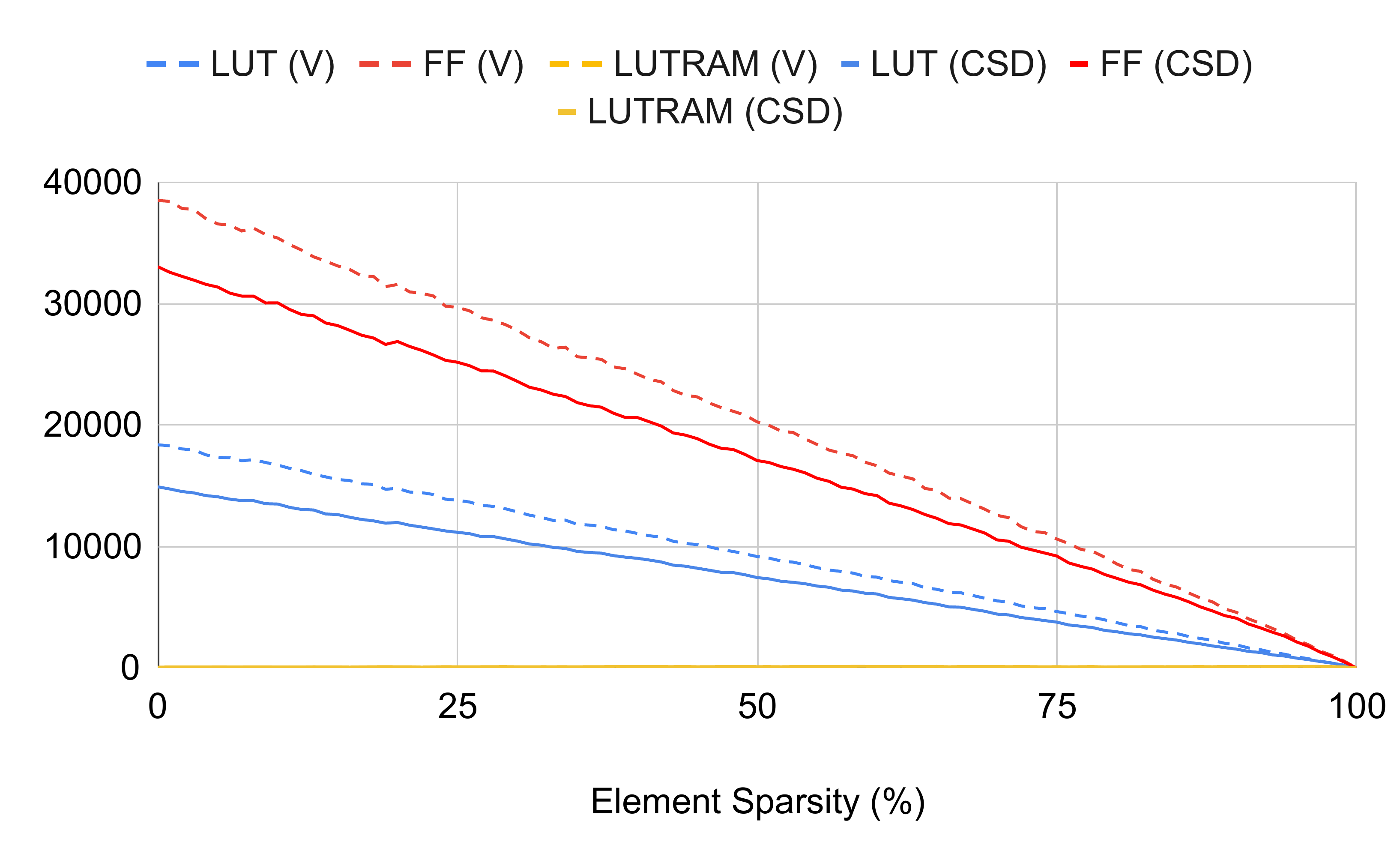}
  \caption{CSD Resource Utilization for 64x64 Element-Sparse matrices}
  \label{fig:csd_es}
\end{figure}

Figure \ref{fig:csd_es} shows the resource utilization when CSD is applied to random matrices of varying element-sparsity. Note that the x-axis is element-sparsity and not bit-sparsity. 

CSD results are strictly better than the naive implementation. In this experiment, all the numbers are uniform random. CSD applies equally to them and reduces the hardware by ~17\% for any level of element-sparsity. In other words, CSD always makes numbers more bit-sparse, which reduces the cost of our architecture. We would expect these savings to improve for larger weight bitwidths.

\section{Large Scale Design Results}
\label{sec:physical}

The prior sections looked at the logic synthesis results of small matrices. In this section, we show the scaling of these techniques to much larger matrices. Our experiments in this section use square matrices with dimensions of 512 and 1024, with 8-bit signed weights. We consider element sparsity from 40\% to 98\%. We use a split matrix to deal with the signed weights, and the split matrices are generated by either splitting positive and negative terms (PN) or by using the CSD technique previously described. The results of these two matrices are fed to an array of final bit-serial subtractors to complete the signed matrix implementation. We "wrap" the matrix multiplier with a small design that feeds inputs from an SRAM, and captures results in that same SRAM. This design wrapper only adds a few extra LUTs and registers. We run synthesis and place and route on these designs and set a timing constraint of 450MHz. Our target FPGA is the Xilinx XCVU13P \cite{ultrascale-plus}, which is a 16nm device containing four chiplets in the package (called Super Logic Regions or SLRs). This device has a capacity of 1.7M 6-input LUTs and 3.4M logic flip-flops.

\begin{figure}[h]
  \centering
  \includegraphics[width=\linewidth, scale=0.9]{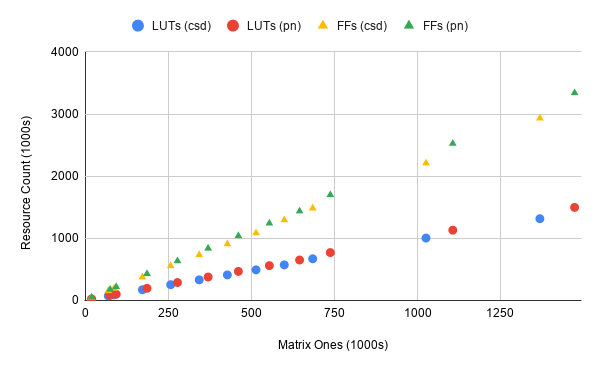}
  \caption{Large Scale Area Results: The very strong linear relationship between matrix ones and FPGA resources is obvious.}
  \label{fig:largescale-area}
\end{figure}

The LUT and register counts as a function of ones in the matrix is plotted in Figure \ref{fig:largescale-area}. The PN and CSD points are based on the identical original signed matrix. The trend lines show the strong correspondence with the resource counts and the number of ones. LUTs are essentially equivalent to the number of ones, and there are two registers per LUT. CSD reduces both the number of ones in the matrix and the resulting resource counts.

\begin{figure}[h]
  \centering
  \includegraphics[width=\linewidth, scale=0.9]{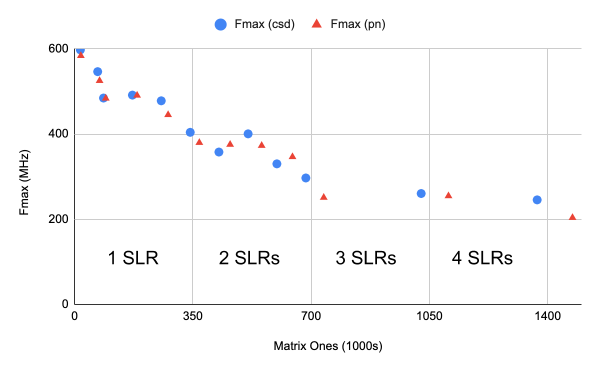}
  \caption{Large Scale Frequency Results: Increasing fanout of the first stage and the spanning of multiple chiplets increase the critical path delay and decrease maximum frequency.}
  \label{fig:largescale-fmax}
\end{figure}

The process of FPGA design does not always meet the specified timing constraint. Figure \ref{fig:largescale-fmax} shows the achieved frequency (Fmax) of these designs after placement and routing. All the paths within these designs have at most one LUT between flops, which means that the frequency is primarily a result of the interconnect delays between LUTs and flops. Other components, such as the SRAMs have a maximum frequency that exceeds 600MHz. There is some noise in these results, but the trends are that bigger matrices run slower. As the matrices get bigger, two mechanisms impact the routing delay:

\begin{itemize}
    \item The initial layer has a large fanout, approximately corresponding to the dimension times the sparsity. Nets that have a fanout of 100s can have delays of several nanoseconds, which becomes the critical path.
    \item Nets cross the chiplet boundaries, and those routes have significantly slower propagation delays.
\end{itemize} 

Each of the four SLRs within the FPGA have a maximum capacity of 425k LUTs. After about 80\% of LUTs are used the tools can struggle. Figure \ref{fig:largescale-fmax} has tick marks illustrating the 82\% thresholds of SLR capacity. Within one SLR, the frequencies range from 597MHz to 445MHz. Designs requiring 2 SLRs range from 296MHz to 400MHz. Matrices bigger than 2 SLRs seem relatively consistent between 225MHz and 250MHz. Both the fanout and chiplet crossing problems could be addressed by adding registers to perform the fanout and chiplet crossings in multiple cycles. These optimizations are not represented here. 

\begin{figure}[h]
  \centering
  \includegraphics[width=\linewidth, scale=0.9]{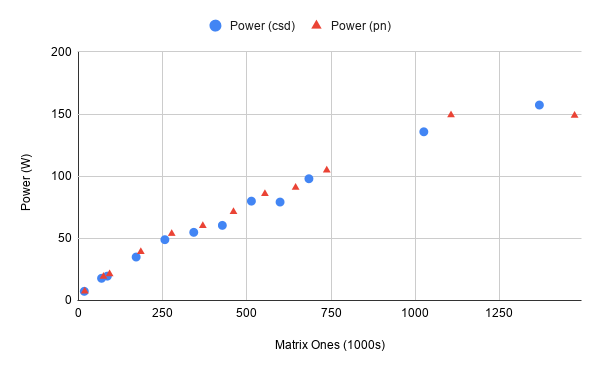}
  \caption{Large Scale Power Results: Note the sublinear increase due to the decreasing acheivable frequency. Under medium cooling assumptions, this FPGA has a limit of about 150W.}
  \label{fig:largescale-pwr}
\end{figure}

The other limit to the frequency is power. Figure \ref{fig:largescale-pwr} plots the estimated total power consumption of this device scaled to run at the maximum achievable frequency. These results were obtained from the Vivado tool based on the default assumptions about switching activity. Under medium settings for airflow and heatsink, the thermal power limit of this FPGA is approximately 150W, which we approach at high dimension and low sparsity.

\section{Evaluation}
\label{sec:eval}
To evaluate our design, we compare it against state-of-the-art sparse multiplication libraries on GPU, and also against a recent sparse DNN accelerator.

\subsection{GPU Benchmarks}
We compare our results against the Volta V100 GPU from NVIDIA. The V100 is built in 12nm technology, whereas the UltraScale+ FPGA is 16nm. The more recent Ampere Architecture from NVIDIA includes hardware support for sparsity up to 50\%, but is unfortunately unavailable to us for this study. We use two libraries for the GPU benchmarks: cuSPARSE \cite{cusparse}, and a recently published optimized kernel \cite{gale2020sparse}. Neither of these libraries support integer arithmetic, so we are using FP16 as a best-case proxy. We run the FPGA at the maximum frequency achieved after place-and-route, which can be seen in Figures \ref{fig:largescale-fmax} and \ref{fig:largescale-pwr}. 

For both of these configurations, we initialize a random weight matrix using the same scheme as described previously. We transfer this to the device and allow the given library to perform any necessary optimizations before timing. Then, we randomly initialize a dense vector, and repeatedly multiply this vector by the matrix so that all caches and memory systems are warm. We do this for 1000 iterations and take the latency to be the mean iteration time. In this case, the latency is measured from the devices memory, through the arithmetic, and back to memory - which is identical to our FPGA implementation.  

\begin{figure}[h]
  \centering
  \includegraphics[width=\linewidth, scale=0.9]{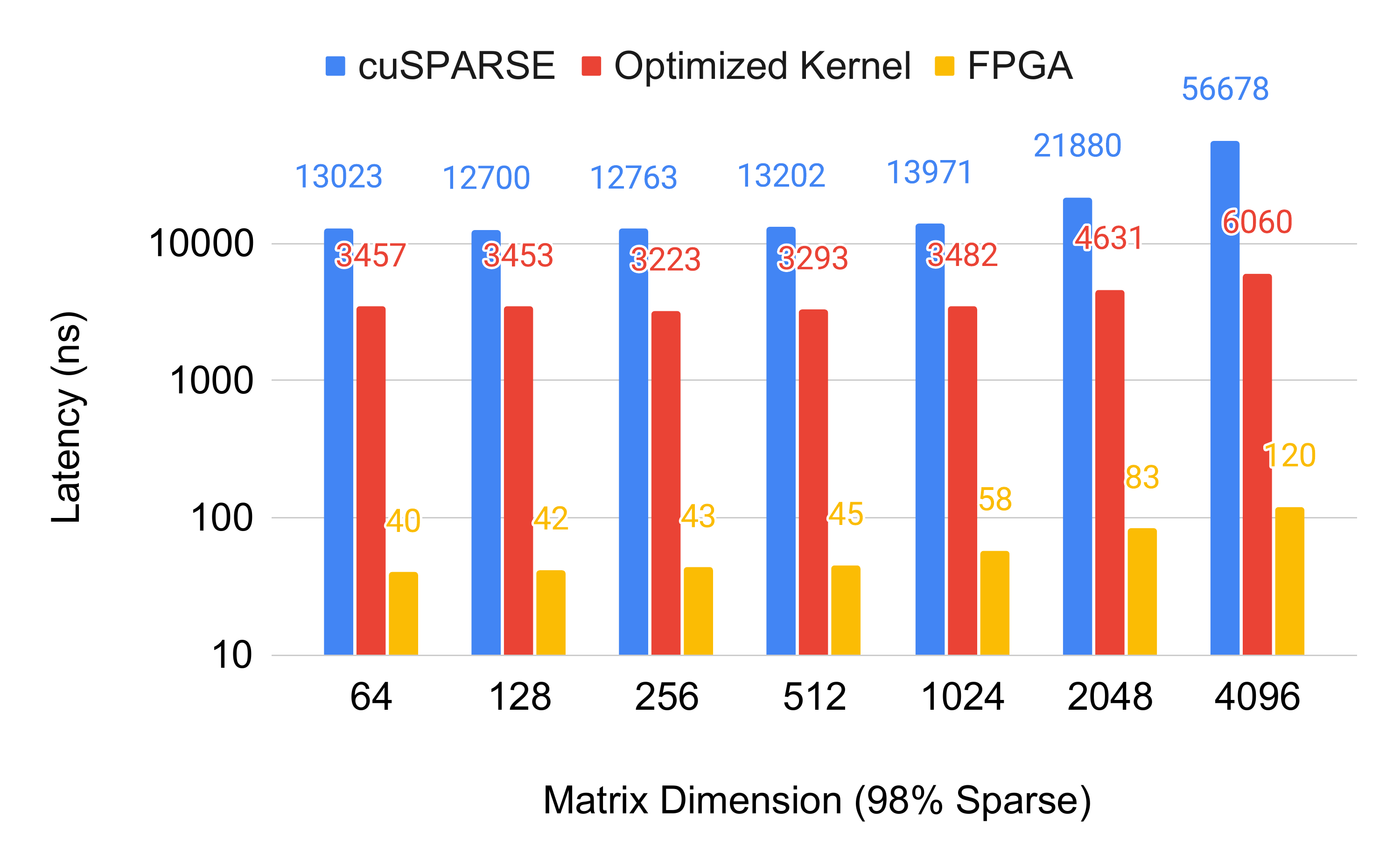}
  \caption{Latency (ns) of varying 98\% element sparse matrices}
  \label{fig:bench_dim_lat}
\end{figure}

\begin{figure}[h]
  \centering
  \includegraphics[width=\linewidth]{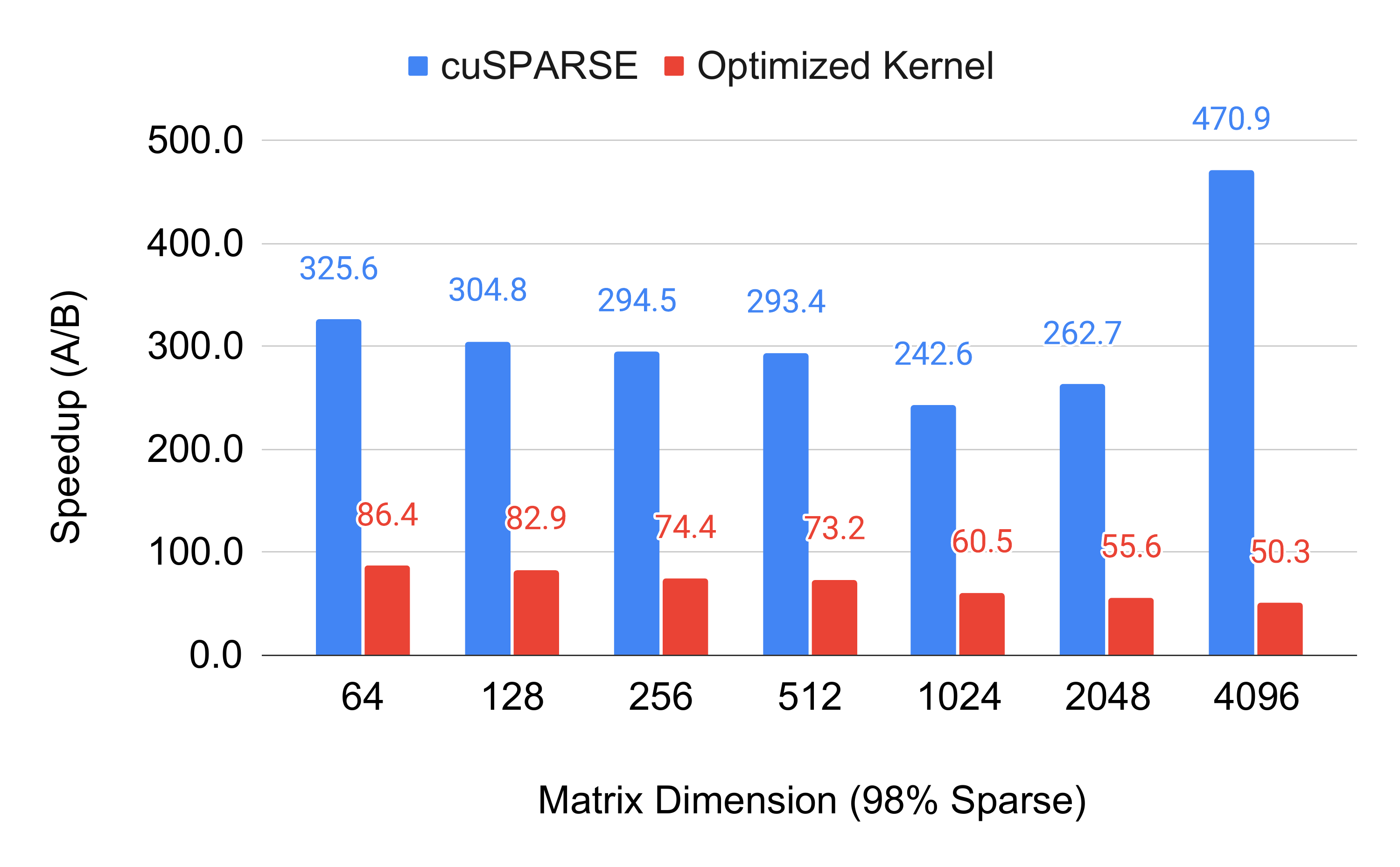}
  \caption{Speedup of varying 98\% element sparse matrices}
  \label{fig:bench_dim_su}
\end{figure}

\paragraph{Sweeping dimension} Our first experiment demonstrates latency performance as a function of array dimensions at 98\% element sparsity. We sweep the matrix dimension from 64x64 to 4096x4096 in powers of two. The results are show in Figure \ref{fig:bench_dim_lat} with actual latency numbers and in Figure \ref{fig:bench_dim_su} as comparative speedup. The "Optimized Kernel" is the library in \cite{gale2020sparse}.

Notice that in all cases, our FPGA latency is less than 120ns, whereas the GPU cannot break the 1$\mu$s barrier. When the matrix size is less than 512x512, the GPU performance is nearly constant. This indicates the GPU is underutilized and therefore  latency bound. GPUs possess incredible throughput using techniques like thread-level parallelism, memory latency hiding, double-buffering, and others. However, all these techniques require the GPU to spawn many more threads than the arithmetic can handle, swapping threads in and out to maximize utilization. In the low-latency regime, these techniques introduce overhead which cannot be overcome, even with the Optimized Kernel. Our solution configures the hardware to exactly support the matrix in question, leading to astounding latency. When the GPU is latency-bound, its latency does not increase for larger matrices. Our solution pays a constant cost to stream the inputs and outputs, but the number of cycles spent reducing partial sum increases logarithmically with matrix size. Furthermore, increasing the matrix size decreases our highest achievable clock frequency. In the GPU's latency-bound regime, we see our speedup fall from 86x to 60x. However, at 1024x1024, the GPU is utilized and is no longer latency-bound, so it begins to see linear scaling with respect to matrix size. Because our solution scales logarithmically in cycles, we see our speedup leveling off at 50x due to the slower clock.

\begin{figure}[h]
  \centering
  \includegraphics[width=\linewidth]{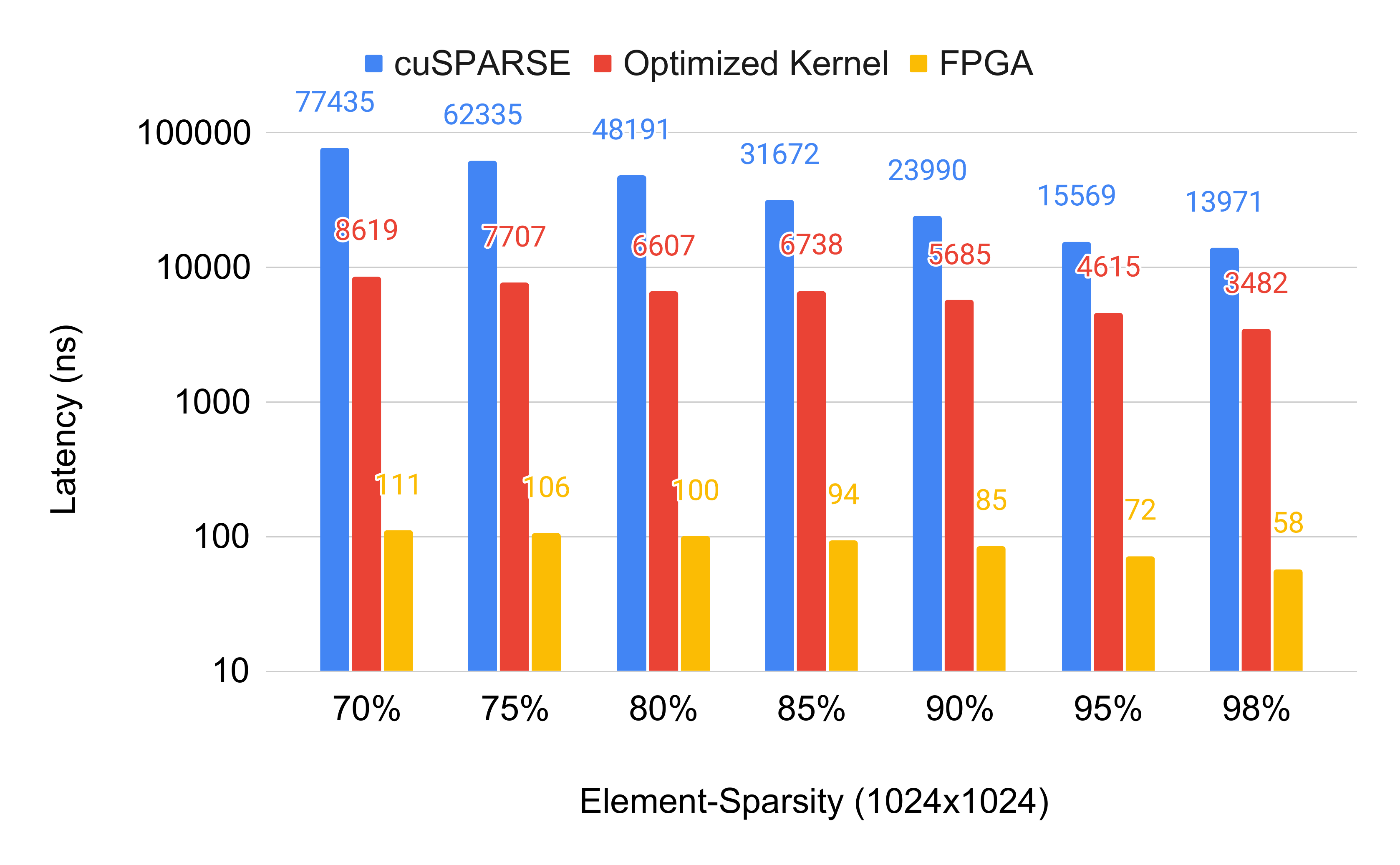}
  \caption{Latency of varying sparsity}
  \label{fig:bench_spar_lat}
\end{figure}

\begin{figure}[h]
  \centering
  \includegraphics[width=\linewidth]{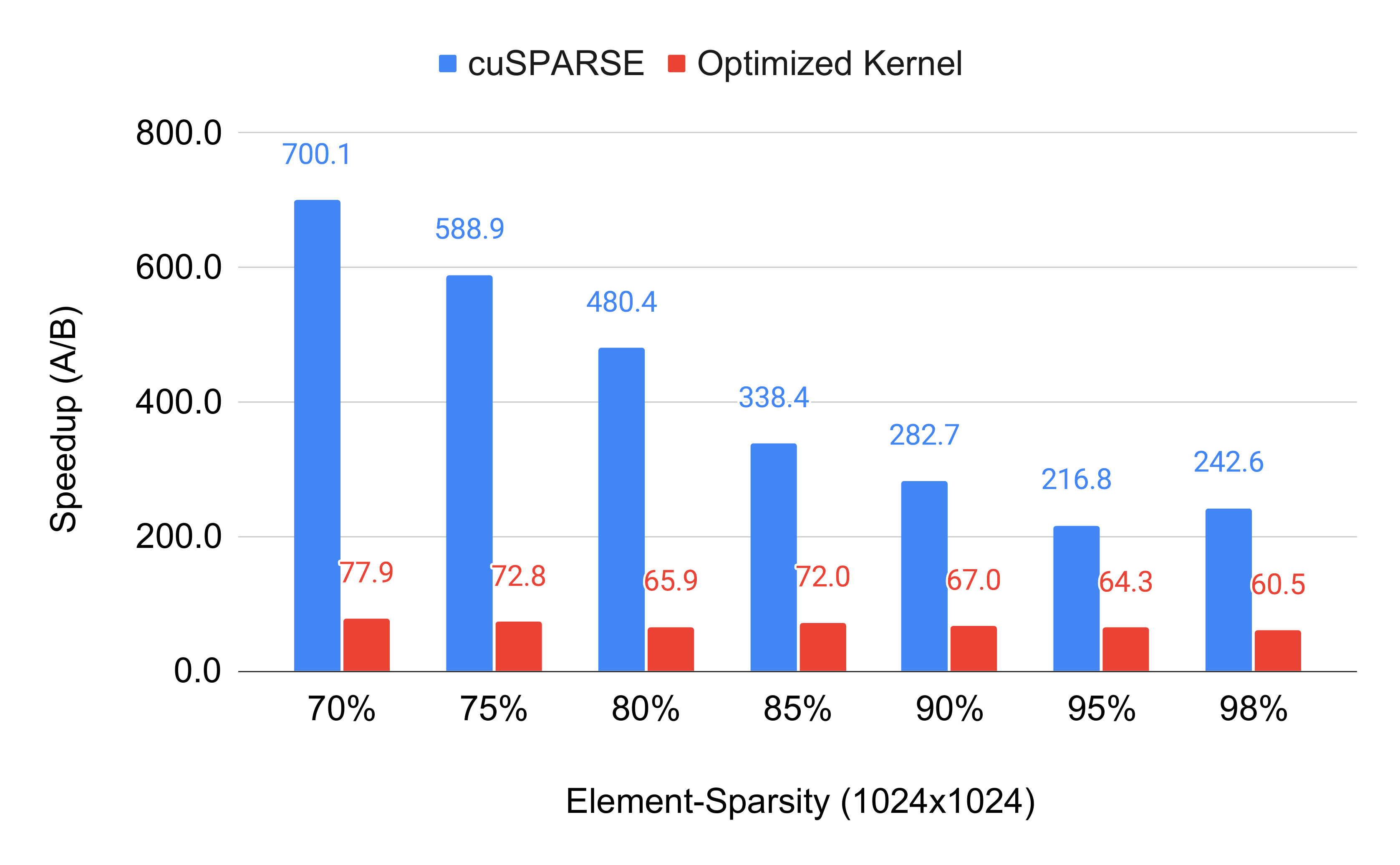}
  \caption{Speedup of varying sparsity}
  \label{fig:bench_spar_su}
\end{figure}

\paragraph{Sweeping sparsity} Our next experiment considers the average latency as a function of matrix sparsity at a fixed dimension of 1024x1024. The results are show in Figures \ref{fig:bench_spar_lat} and \ref{fig:bench_spar_su} with average latency and speedup, respectively. Element sparsity is swept from 70\% to 98\%. Recall that our solution's latency in cycles does not depend on sparsity, but we can clock the design faster as sparsity increases. The GPU sees decreasing latency with increasing sparsity because it needs to operate on fewer elements and the cost of indexing is amortized. For cuSPARSE, increasing sparsity from 70\% to 85\% sees large reductions in latency as the library reduces the amount of compute to be done. The optimized kernel comparatively spends less time indexing and has higher performance at lower sparsity. In both cases, the compute reduction from 70\% to 85\% is effective and sees our speedup go from 77x to 72x. As sparsity increases further, the GPU again becomes underutilized and both the latency and speedup level off, yielding a minimum speedup of 60x.  Again, the GPU is unable to break the 1$\mu$s barrier, whereas our solution stays under 120ns. 

\paragraph{Batching} We have illustrated the advantage of our architecture for latency-sensitive scenarios, which is our targeted case. In this next experiment, we compare throughput of the GPU solution versus our solution. Previous experiments, because of the recurrence, were done with a batch size of 1, which underutilizes the GPU's pipelined resources. In this experiment, we change the GPU's computation to perform matrix-matrix multiplication, where the number of columns in the multiplicand matrix is the "batch-size", borrowing terminology from DNN processing. We again randomly initialize a fixed weight matrix with 95\% sparsity and signed 8-bit integers. We choose 95\% sparsity to give the GPU ample headroom. We sweep the batch size from 1 to 64, reporting the speedup of our solution compared to the GPU. We are still measuring latency, but in the higher batch sizes, the reciprocal of the latency will approach the maximum throughput of the GPU.

\begin{figure}[h]
  \centering
  \includegraphics[width=\linewidth]{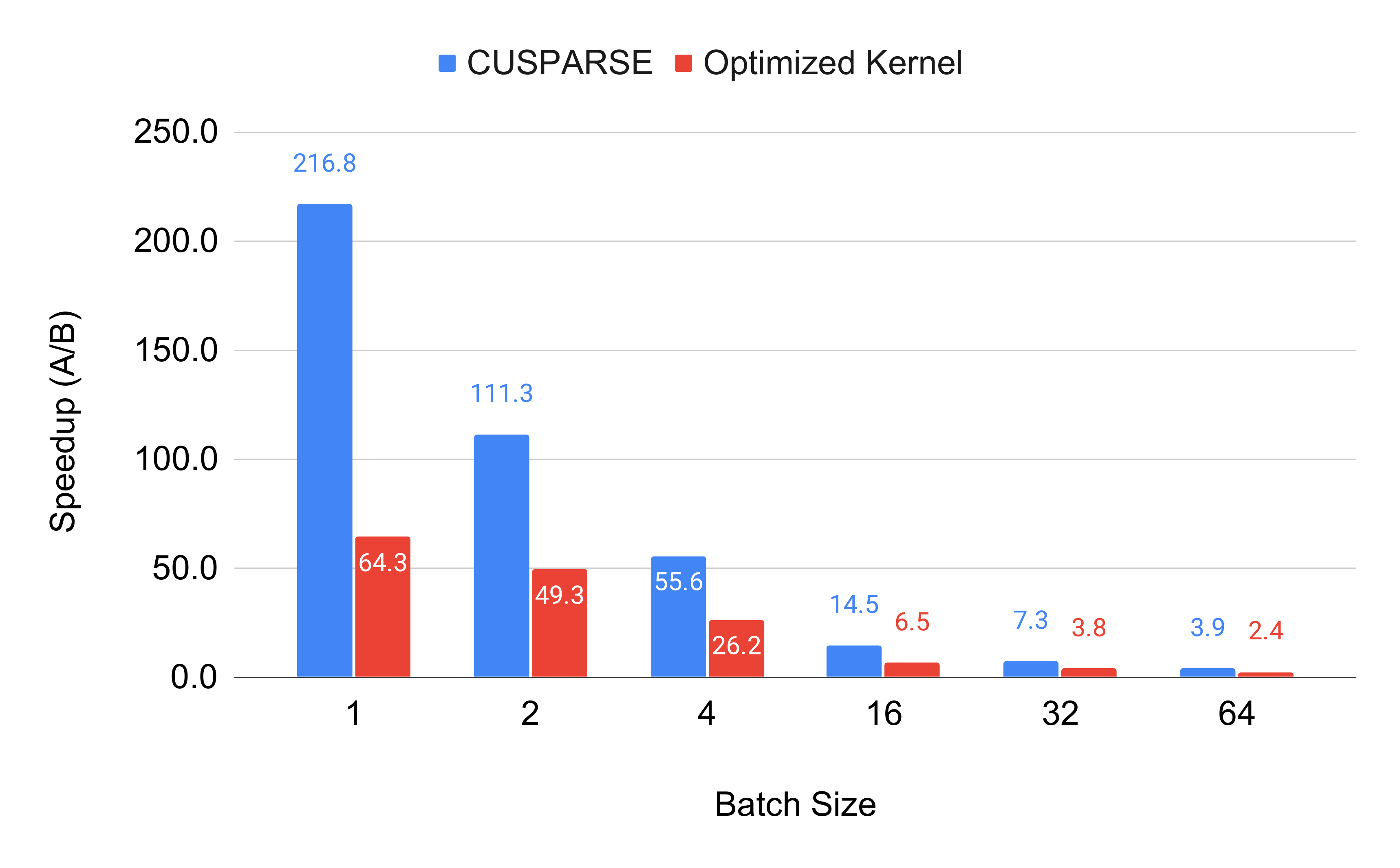}
  \caption{Speedup against V100 for varying batch size (matrix = 1024x1024, 95\% sparse) Batch-size 1 is a comparison of pure latency. As batch size increases, the speedup is comparing achievable throughput}
  \label{fig:bench_batch_1024}
\end{figure}

Figure \ref{fig:bench_batch_1024} presents our results for a 1024x1024 weight matrix, and multiplying a batch\_size x 1024 random dense matrix.

\begin{figure}[h]
  \centering
  \includegraphics[width=\linewidth]{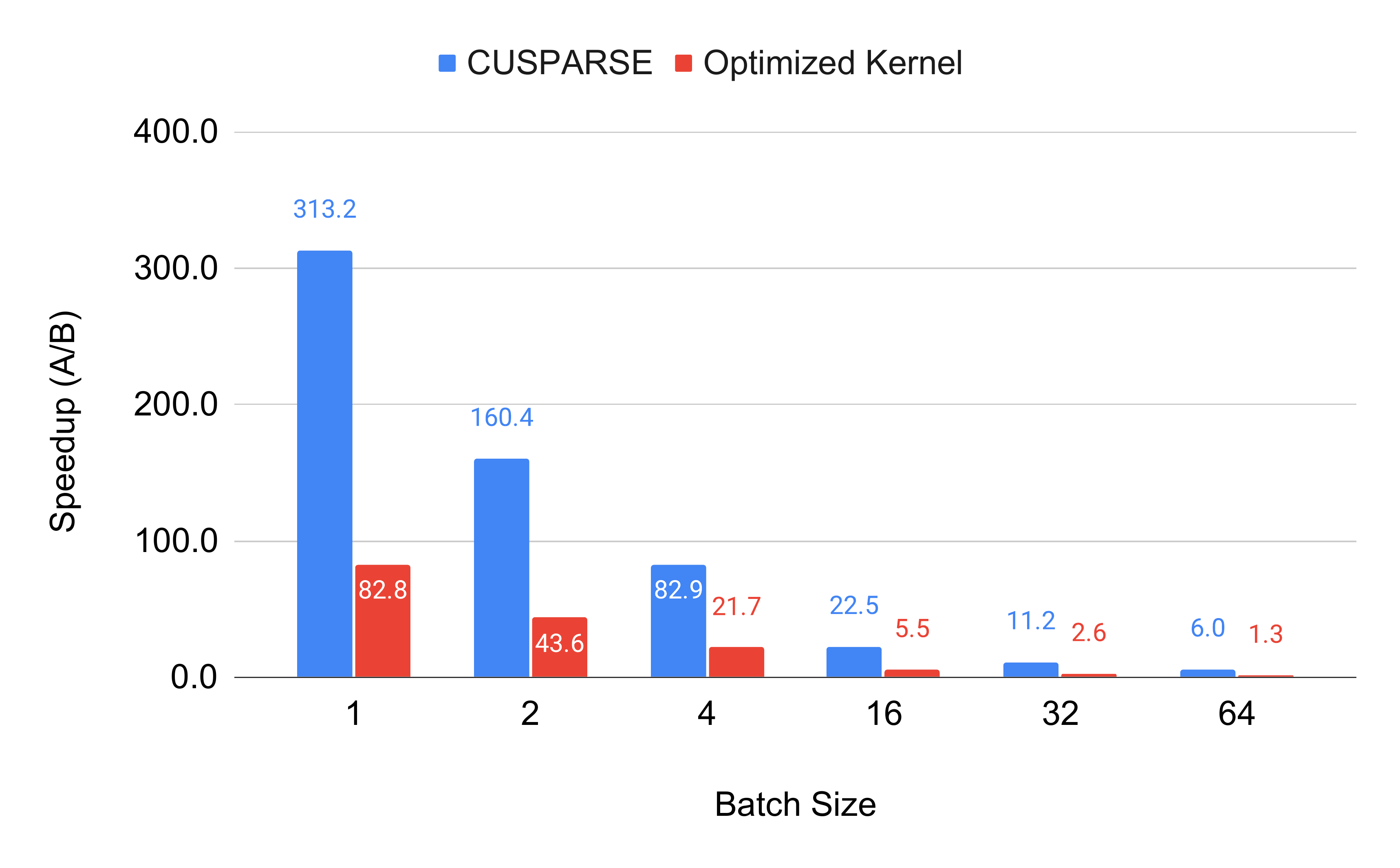}
  \caption{Speedup against V100 for varying batch size (matrix = 64x64, 95\% sparse) Batch-size 1 is a comparison of pure latency. As batch size increases, the speedup is comparing achievable throughput}
  \label{fig:bench_batch_64}
\end{figure}

Figure \ref{fig:bench_batch_64} presents our results for a 64x64 weight matrix, and multiplying a batch\_size x 64 random dense matrix.

As expected, the latency for the GPU solution scales sub-linearly with respect to batch size. As the GPU becomes more utilized, it's able to overlap computation and memory to take full advantage of the massive parallelism. Because our architecture is built for vector products, we have to stream the columns of the input matrix in one-by-one, which yields to linear scaling. Because our solution begins with such a lead, it's able to stream multiple batches before the GPU can reach the same performance. This means our solution is still lower latency at small batch sizes. In the 1024 case, our solution is still marginally better due to the GPU already being close to utilized with the large matrix. With 64x64, the GPU has more computational intensity to fill before it becomes utilized.

The TDP of our FPGA is 150W, while the V100 is 300W, but these numbers are not reliable for accurate run-time comparisons. Given the disparity between the platforms, it is very difficult to accurately compare power consumption between these two solutions. There is a difference in the process technology. The FPGA and GPU also have a different set of peripherals associated with them, that may or may not be active. The dynamic power will be different based on the specific variations of the random weights and activations, which is exacerbated by the int8 vs fp16 difference. Finally, the different boards have different methods to monitor power consumption on their different supply rails. However, we believe that the efficiency gains shown here are due to fundamental computational simplification, and it would be reasonable to assume that the dynamic power would be correspondingly lower. 

\subsection{DNN Accelerator Benchmarks}
We also consider the state-of-the-art in Sparse DNN acceleration: SIGMA \cite{sigma}. SIGMA similarly relies on an input broadcast and reduction tree to perform matrix-multiplication. We reached out to the authors and were given their cycle-accurate simulator. In their paper, they design SIGMA as a 128x128 array of fp16 processing elements (PEs) clocked at 500MHz. To approximate process technology node differences and the change to int8 from fp16, we assume that SIGMA can be clocked at 1GHz. We run SIGMA with the weight matrix stationary and stream the input in to minimize latency.

\begin{figure}[h]
  \centering
  \includegraphics[width=\linewidth]{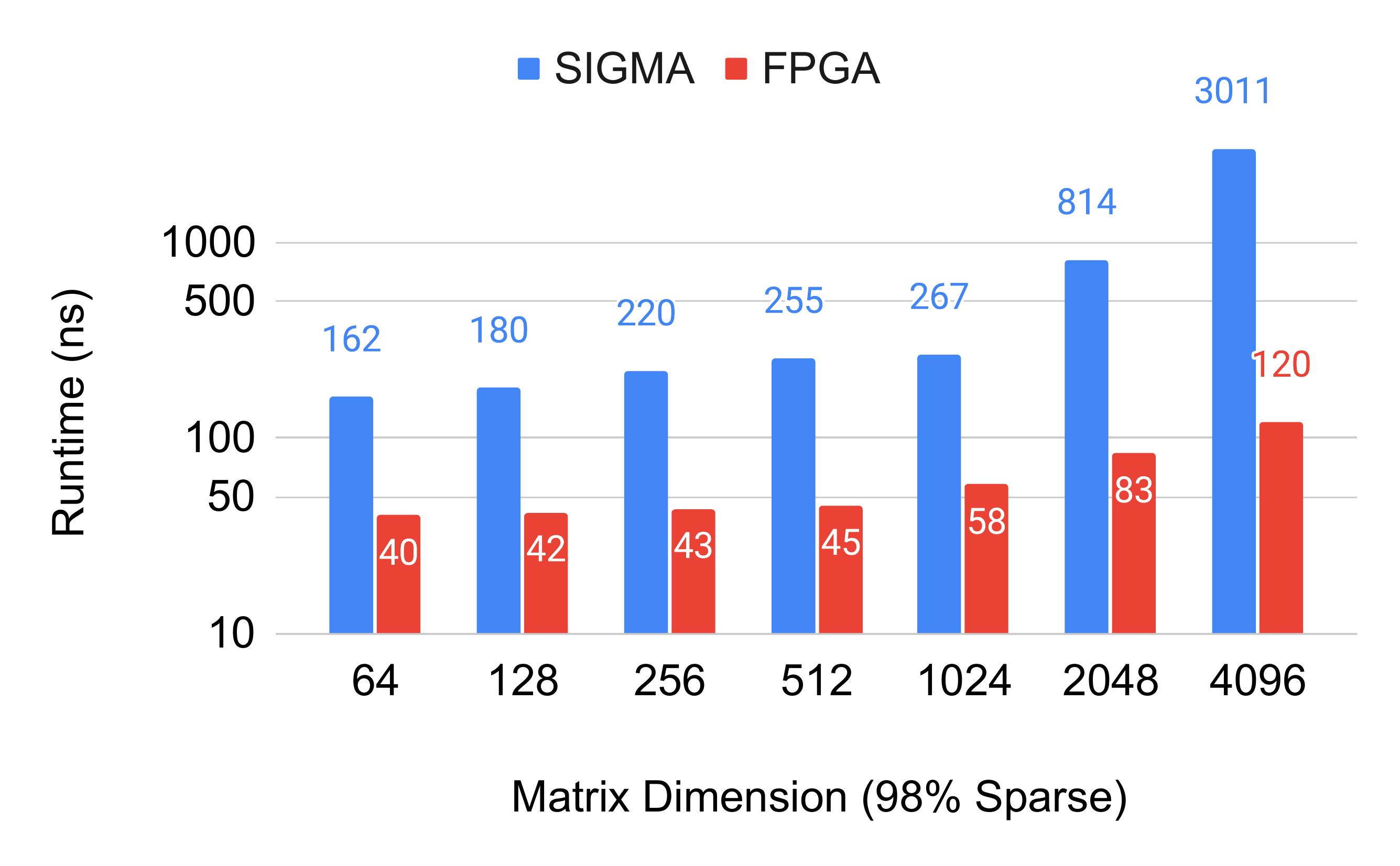}
  \caption{Latency of FPGA and SIGMA for varying dimension (98\% sparse)}
  \label{fig:bench_sigma_dim_lat}
\end{figure}

\begin{figure}[h]
  \centering
  \includegraphics[width=\linewidth]{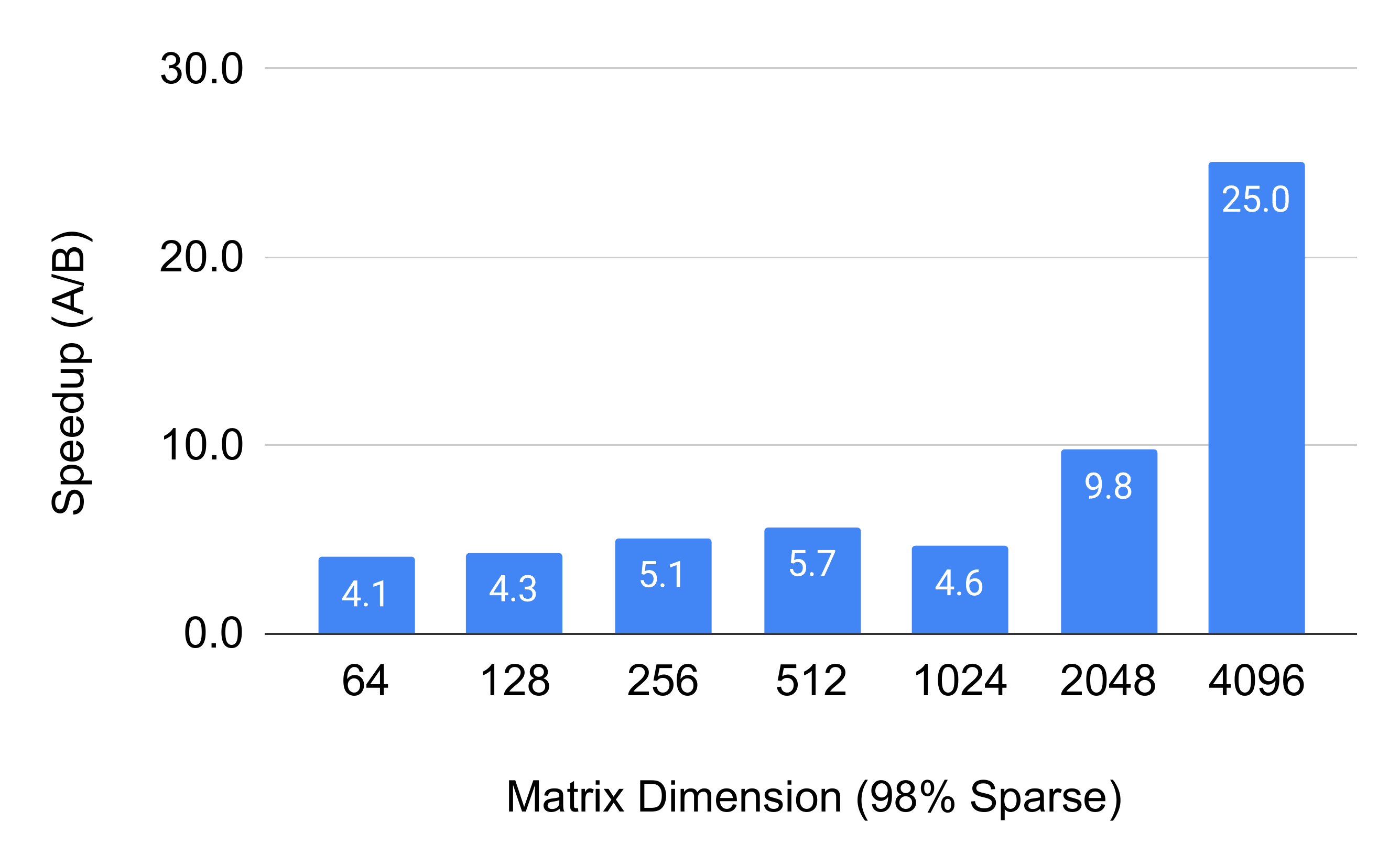}
  \caption{Speedup against SIGMA for varying dimension (98\% sparse)}
  \label{fig:bench_sigma_dim_su}
\end{figure}

\paragraph{Sweeping dimension} We repeat a similar suite of experiments as before. We sweep the dimension of 98\% element-sparse matrices. Figures \ref{fig:bench_sigma_dim_lat} and \ref{fig:bench_sigma_dim_su} show the latency and speedup respectively. 

For small dimensions, SIGMA does report nanosecond-scale latency due to its input broadcast and reduction tree. Furthermore, small, sparse matrices easily fit into the PE grid, so there is little overhead from tiling. However, after 1024x1024, the elements no longer fit in the PE grid and the computation must be tiled. This invokes extra SRAM use and transitions SIGMA into the memory-bound region, where it sees linear scaling. This yields a 4.1x speedup for our solution in the worst case, but we quickly gain a 25x advantage as the matrix size increases. 

\begin{figure}[h]
  \centering
  \includegraphics[width=\linewidth]{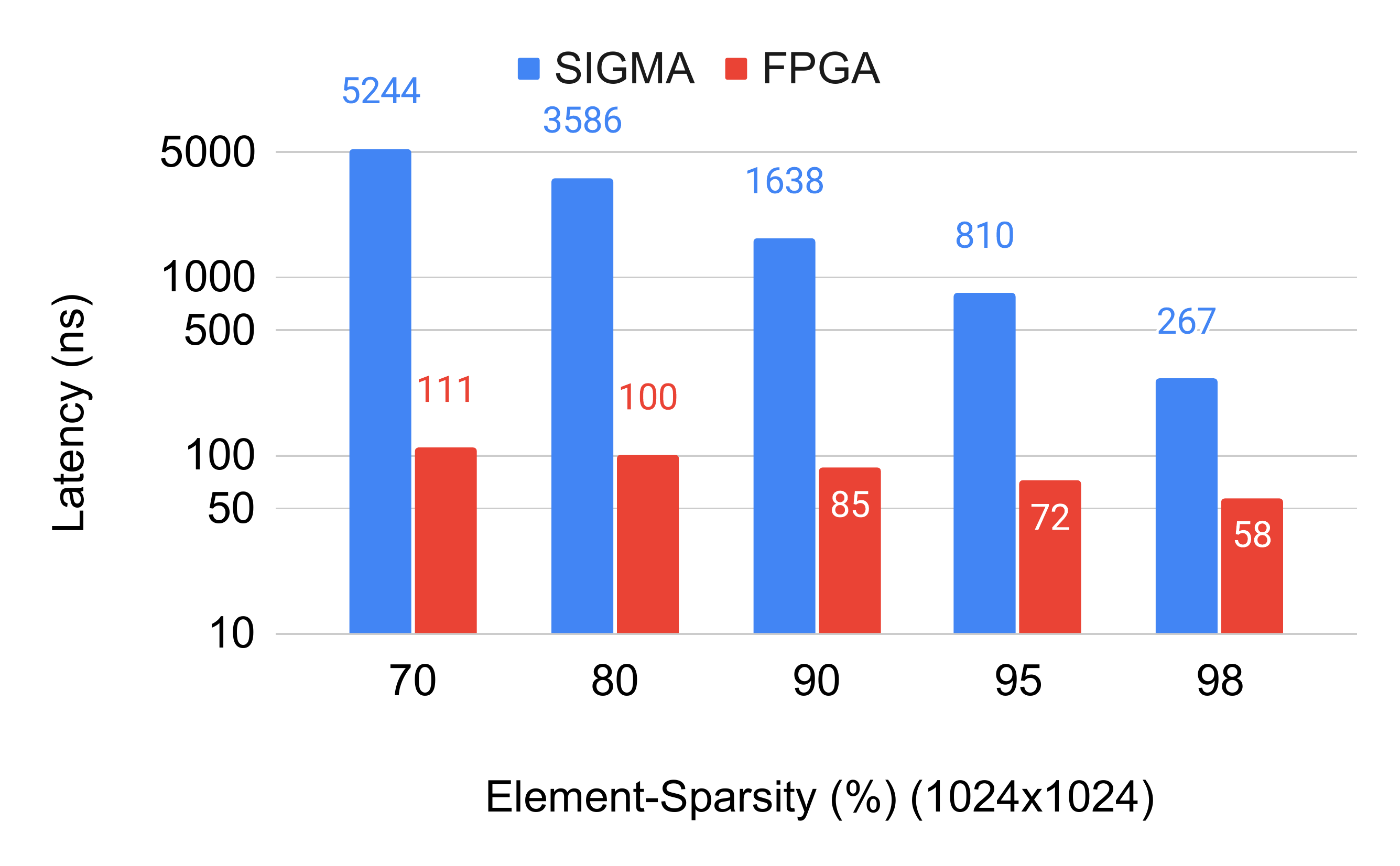}
  \caption{Latency of FPGA and SIGMA for varying sparsity (1024x1024)}
  \label{fig:bench_sigma_spar_lat}
\end{figure}

\begin{figure}[h]
  \centering
  \includegraphics[width=\linewidth]{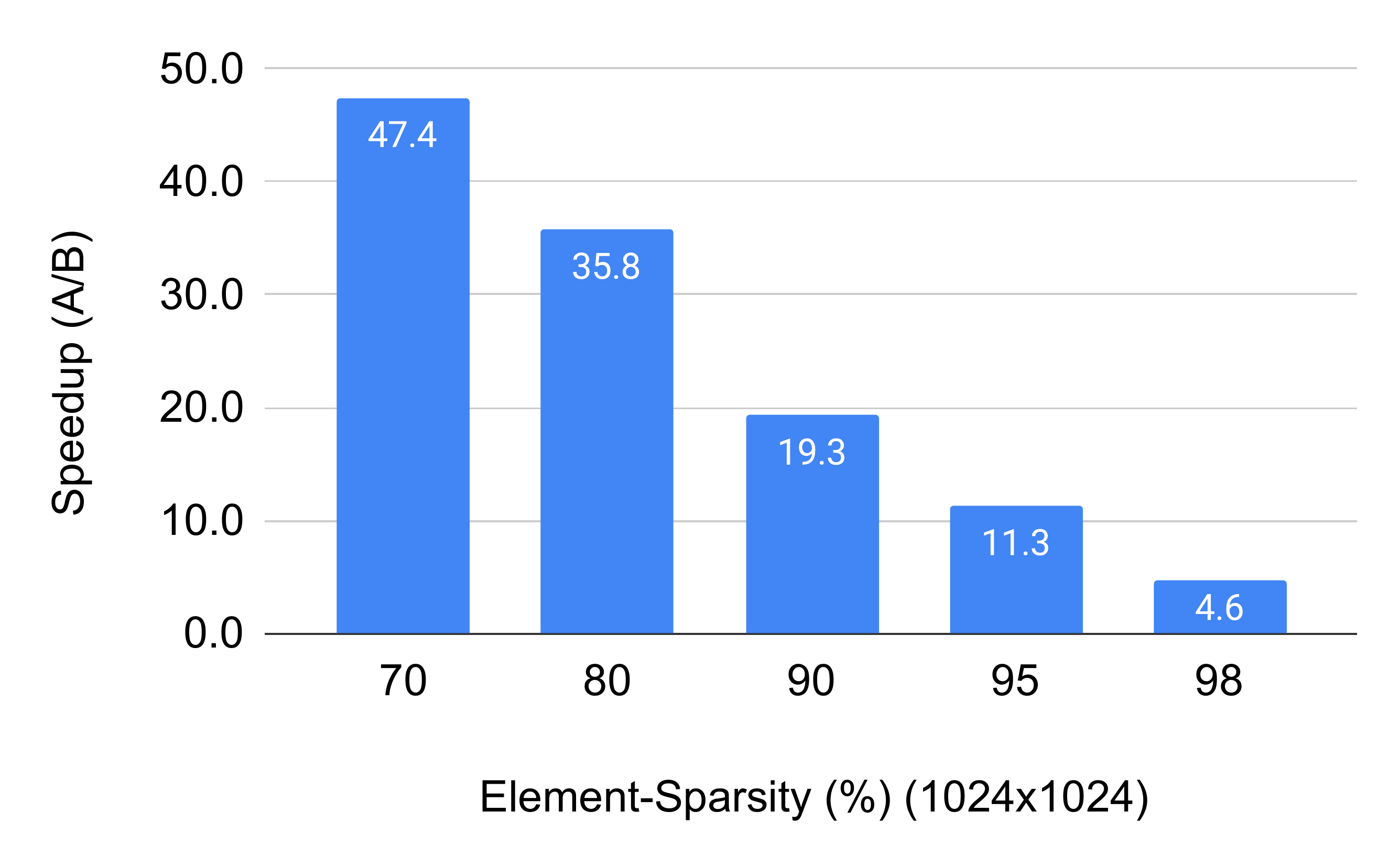}
  \caption{Speedup against SIGMA for varying sparsity (1024x1024)}
  \label{fig:bench_sigma_spar_su}
\end{figure}

\paragraph{Sweeping sparsity} Moving on, Figures \ref{fig:bench_sigma_spar_lat} and \ref{fig:bench_sigma_spar_su} report the latency and speedup of a 1024x1024 matrix, sweeping element-sparsity.

The advantage of SIGMA is that it only maps non-zero weight and activation pairs to PEs. If this union can fit into the PE grid, no tiling is done. As we expect, the size of this set is directly related to the sparsity, and SIGMA sees huge latency improvements as sparsity increases, taking it into the nanosecond regime. However, even 90\% sparsity and below is enough to push it back into the microsecond regime, which yields a large advantage to our design.

\begin{figure}[h]
  \centering
  \includegraphics[width=\linewidth]{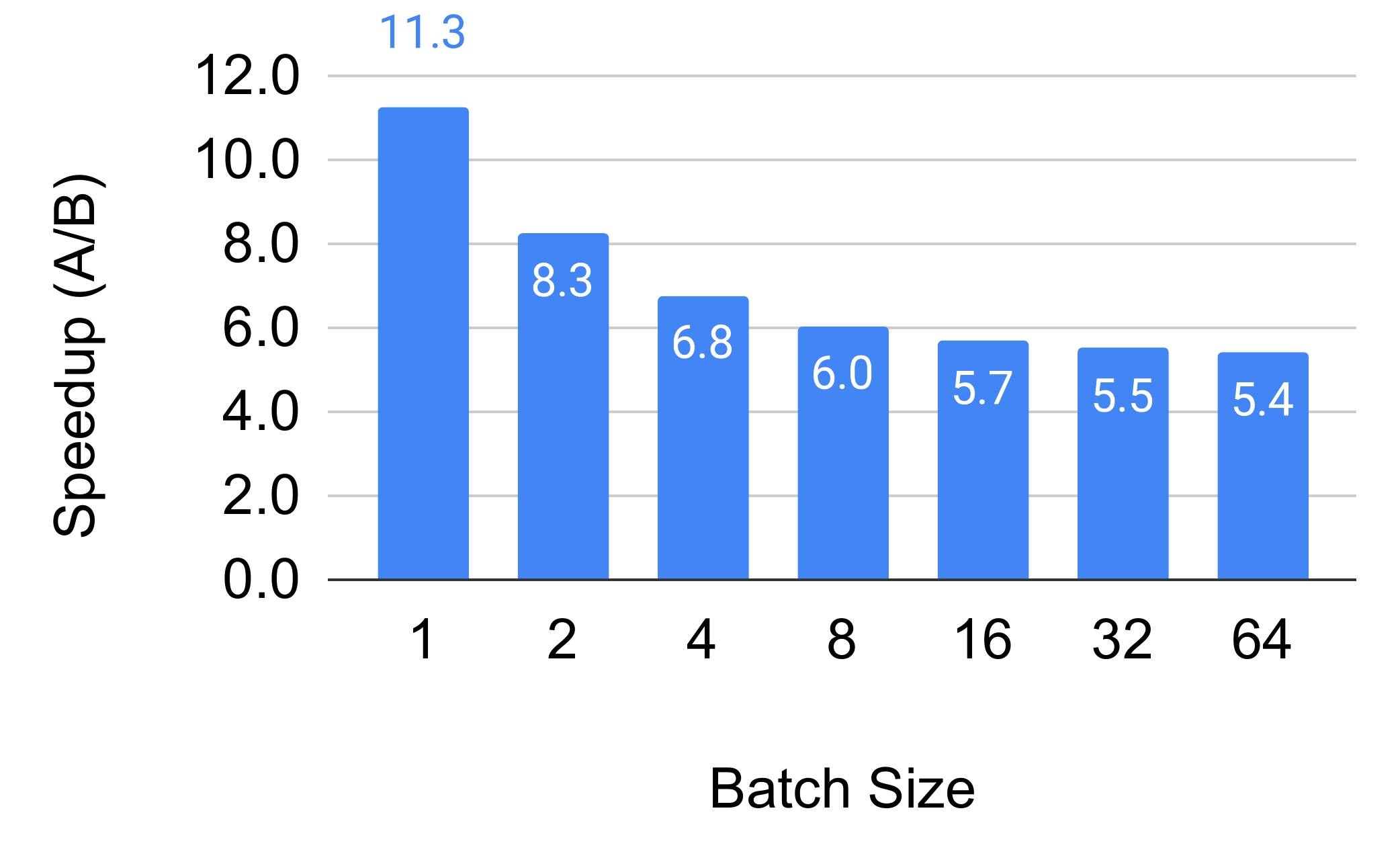}
  \caption{Speedup against SIGMA of varying batch-size (1024x1024, 95\%) Batch-size 1 is a comparison of pure latency. As batch size increases, the speedup is comparing achievable throughput}
  \label{fig:bench_sigma_batch_1024}
\end{figure}

\paragraph{Batching} Finally, we compare matrix-matrix multiplication between our solution and SIGMA. We repeat the same batching test and the result is shown in Figure  \ref{fig:bench_sigma_batch_1024}.

Because SIGMA is able to fill its PE array with sparse entries at small batches, it doesn't have as much to gain as a GPU when transitioning from vector to matrix products. At batch-size 2, SIGMA does find opportunity to utilize more PEs and our advantage decreases. However, batch-size 4 and beyond quickly see SIGMA in the memory-bound region again, which causes the speedup to saturate at 5.4x.

\section{Discussion and Future Work}
\label{sec:future}

For the given use case, our design has two major limitations: 1) the fanout of the input broadcast saturates the interconnect resources of the FPGA and limits frequency. 2) we are bound by the number of 6-input LUTs in the FPGA, which limits the number of set bits we have in the weight matrix. Creating a CGRA architecture could solve both of these issues. We could employ a broadcast-friendly pipelined interconnect network for the inputs, similar to the Benes network in SIGMA. Furthermore, a 6-input LUT is made using 64 SRAM bits of 6 transistors each, with 64 MUX T-gates of 2 transistors each, which yields a total of 512 transistors for every LUT. A full-adder uses 16 or fewer transistors \cite{full-adder-16t}, which is 1/32 the cost. A CGRA implementation of our design would see a grid of full-adders and flip-flops, with a flexible tree-like interconnect to perform partial sums and broadcast interconnect for the input. This approach would allow for higher compute density at higher frequencies.

Even with these optimizations, there may be instances where the compute matrix cannot entirely fit in hardware and must be tiled similar to DNN accelerators. To amortize the cost of loading and moving weights, DNN inference accelerators often employ batching to perform the multiplication of multiple vectors with the same matrix, which saves power by reducing the movement of matrix weights. Our approach eliminates the movement of matrix weights by programming them into an interconnect matrix and takes further advantage by propagating the constants within the matrix. The time to modify the interconnect matrix of the FPGA is on the order of 200ms, which limits its practicality in moving weights during runtime.

However, the feed-forward topology of this network allows for the approach of pipeline reconfiguration \cite{piperench99}. Pipeline configuration is the ability to configure hardware cycle-by-cycle as the pipeline fills. This concept would allow a reconfigurable platform with nearly zero configuration overhead time to change the matrix weights. Pipeline configuration is not supported in conventional FPGAs, but we could support it in a custom CGRA architecture. As the partial sums travel down the tree, the levels above the current accumulation can be recongifured as their state is no longer needed. One can think of "waves" of configuration travelling down the tree, where some parts of the tree are reconfiguring and some parts are computing. This is analogous to double-buffering in DNN processing.

We plan to explore these optimizations in future work.

\section{Conclusion}
We present an architecture for performing integer matrix-vector multiplication. The cost and power of our implementations scale exactly with the number of non-zero terms in the matrix. Bit serial implementations allow for us to support matrices with up to 1.5 million ones, as large as 1024x1024 eight-bit matrix at a sparsity of 60\%. These products can be produced with nanosecond-scale latencies, which is 2-3 orders of magnitude faster than a GPU, and 4-47x faster than a sparse accelerator. Throughput is also competitive, especially for large matrices. Given an FPGA's slow configuration time, these techniques are primarily suitable to computations where the matrix is essentially fixed, and where low latency is critical, such as reservoir computing. However, a customized programmable device for this approach could pipeline the configuration, thus effectively hiding it, and enable this approach to work for dynamic sparse matrices.

\section*{Acknowledgements}
We would like to thank Peter Matheu for the extensive technical discussion and help with the SystemVerilog implementation. We would also like to thank Amir Yazdanbakhsh and Dan Zhang for paper review, and Guillermo Maturana for code review. This work was completed as an internship project.

\bibliographystyle{IEEEtranS}
\bibliography{acmart}

\begin{thebibliography}{10}
\providecommand{\url}[1]{#1}
\csname url@samestyle\endcsname
\providecommand{\newblock}{\relax}
\providecommand{\bibinfo}[2]{#2}
\providecommand{\BIBentrySTDinterwordspacing}{\spaceskip=0pt\relax}
\providecommand{\BIBentryALTinterwordstretchfactor}{4}
\providecommand{\BIBentryALTinterwordspacing}{\spaceskip=\fontdimen2\font plus
\BIBentryALTinterwordstretchfactor\fontdimen3\font minus
  \fontdimen4\font\relax}
\providecommand{\BIBforeignlanguage}[2]{{%
\expandafter\ifx\csname l@#1\endcsname\relax
\typeout{** WARNING: IEEEtranS.bst: No hyphenation pattern has been}%
\typeout{** loaded for the language `#1'. Using the pattern for}%
\typeout{** the default language instead.}%
\else
\language=\csname l@#1\endcsname
\fi
#2}}
\providecommand{\BIBdecl}{\relax}
\BIBdecl

\bibitem{numenta2019}
\BIBentryALTinterwordspacing
S.~Ahmad and L.~Scheinkman, ``How can we be so dense? the benefits of using
  highly sparse representations,'' \emph{CoRR}, vol. abs/1903.11257, 2019.
  [Online]. Available: \url{http://arxiv.org/abs/1903.11257}
\BIBentrySTDinterwordspacing

\bibitem{bit-pragmatic}
\BIBentryALTinterwordspacing
J.~Albericio, P.~Judd, A.~D. Lascorz, S.~Sharify, and A.~Moshovos,
  ``Bit-pragmatic deep neural network computing,'' \emph{CoRR}, vol.
  abs/1610.06920, 2016. [Online]. Available:
  \url{http://arxiv.org/abs/1610.06920}
\BIBentrySTDinterwordspacing

\bibitem{antonik2015fpga}
P.~Antonik, A.~Smerieri, F.~Duport, M.~Haelterman, and S.~Massar, ``Fpga
  implementation of reservoir computing with online learning,'' in \emph{24th
  Belgian-Dutch Conference on Machine Learning}, 2015.

\bibitem{signed_digit}
A.~{Avizienis}, ``Signed-digit number representations for fast parallel
  arithmetic,'' \emph{IRE Transactions on Electronic Computers}, vol. EC-10,
  no.~3, pp. 389--400, 1961.

\bibitem{bianchi20}
F.~M. {Bianchi}, S.~{Scardapane}, S.~{Løkse}, and R.~{Jenssen}, ``Reservoir
  computing approaches for representation and classification of multivariate
  time series,'' \emph{IEEE Transactions on Neural Networks and Learning
  Systems}, pp. 1--11, 2020.

\bibitem{stochastic}
\BIBentryALTinterwordspacing
B.~S. David~Verstraeten and D.~Stroobandt, ``Reservoir computing with
  stochastic bitstream neurons,'' \emph{Proc of ProRISC Workshop}, 2005.
  [Online]. Available:
  \url{http://citeseerx.ist.psu.edu/viewdoc/download?doi=10.1.1.60.5025&rep=rep1&type=pdf}
\BIBentrySTDinterwordspacing

\bibitem{full-adder-16t}
A.~{Dubey}, S.~{Akashe}, and S.~{Dubey}, ``A novel high-performance cmos 1 bit
  full-adder cell,'' in \emph{2013 7th International Conference on Intelligent
  Systems and Control (ISCO)}, 2013, pp. 312--315.

\bibitem{lth18}
\BIBentryALTinterwordspacing
J.~Frankle and M.~Carbin, ``The lottery ticket hypothesis: Training pruned
  neural networks,'' \emph{CoRR}, vol. abs/1803.03635, 2018. [Online].
  Available: \url{http://arxiv.org/abs/1803.03635}
\BIBentrySTDinterwordspacing

\bibitem{gale2020sparse}
\BIBentryALTinterwordspacing
T.~Gale, M.~Zaharia, C.~Young, and E.~Elsen, ``Sparse gpu kernels for deep
  learning,'' 2020. [Online]. Available: \url{https://arxiv.org/abs/2006.10901}
\BIBentrySTDinterwordspacing

\bibitem{gallicchio2020sparsity}
\BIBentryALTinterwordspacing
C.~Gallicchio, ``Sparsity in reservoir computing neural networks,'' 2020.
  [Online]. Available: \url{https://arxiv.org/abs/2006.02957}
\BIBentrySTDinterwordspacing

\bibitem{piperench99}
\BIBentryALTinterwordspacing
S.~C. Goldstein, H.~Schmit, M.~Moe, M.~Budiu, S.~Cadambi, R.~R. Taylor, and
  R.~Laufer, ``Piperench: A co/processor for streaming multimedia
  acceleration,'' in \emph{Proceedings of the 26th Annual International
  Symposium on Computer Architecture}, ser. ISCA '99.\hskip 1em plus 0.5em
  minus 0.4em\relax USA: IEEE Computer Society, 1999, p. 28–39. [Online].
  Available: \url{https://doi.org/10.1145/300979.300982}
\BIBentrySTDinterwordspacing

\bibitem{han2016eie}
\BIBentryALTinterwordspacing
S.~Han, X.~Liu, H.~Mao, J.~Pu, A.~Pedram, M.~A. Horowitz, and W.~J. Dally,
  ``Eie: Efficient inference engine on compressed deep neural network,'' 2016.
  [Online]. Available: \url{https://arxiv.org/abs/1602.01528}
\BIBentrySTDinterwordspacing

\bibitem{han2016deep}
\BIBentryALTinterwordspacing
S.~Han, H.~Mao, and W.~J. Dally, ``Deep compression: Compressing deep neural
  networks with pruning, trained quantization and huffman coding,'' 2016.
  [Online]. Available: \url{https://arxiv.org/abs/1510.00149}
\BIBentrySTDinterwordspacing

\bibitem{tpu}
\BIBentryALTinterwordspacing
N.~P. Jouppi, C.~Young, N.~Patil, D.~A. Patterson, G.~Agrawal, R.~Bajwa,
  S.~Bates, S.~Bhatia, N.~Boden, A.~Borchers, R.~Boyle, P.~Cantin, C.~Chao,
  C.~Clark, J.~Coriell, M.~Daley, M.~Dau, J.~Dean, B.~Gelb, T.~V. Ghaemmaghami,
  R.~Gottipati, W.~Gulland, R.~Hagmann, R.~C. Ho, D.~Hogberg, J.~Hu, R.~Hundt,
  D.~Hurt, J.~Ibarz, A.~Jaffey, A.~Jaworski, A.~Kaplan, H.~Khaitan, A.~Koch,
  N.~Kumar, S.~Lacy, J.~Laudon, J.~Law, D.~Le, C.~Leary, Z.~Liu, K.~Lucke,
  A.~Lundin, G.~MacKean, A.~Maggiore, M.~Mahony, K.~Miller, R.~Nagarajan,
  R.~Narayanaswami, R.~Ni, K.~Nix, T.~Norrie, M.~Omernick, N.~Penukonda,
  A.~Phelps, J.~Ross, A.~Salek, E.~Samadiani, C.~Severn, G.~Sizikov,
  M.~Snelham, J.~Souter, D.~Steinberg, A.~Swing, M.~Tan, G.~Thorson, B.~Tian,
  H.~Toma, E.~Tuttle, V.~Vasudevan, R.~Walter, W.~Wang, E.~Wilcox, and D.~H.
  Yoon, ``In-datacenter performance analysis of a tensor processing unit,''
  \emph{CoRR}, vol. abs/1704.04760, 2017. [Online]. Available:
  \url{http://arxiv.org/abs/1704.04760}
\BIBentrySTDinterwordspacing

\bibitem{stripes}
P.~{Judd}, J.~{Albericio}, and A.~{Moshovos}, ``Stripes: Bit-serial deep neural
  network computing,'' \emph{IEEE Computer Architecture Letters}, vol.~16,
  no.~1, pp. 80--83, 2017.

\bibitem{kleyko2020integer}
\BIBentryALTinterwordspacing
D.~Kleyko, E.~P. Frady, M.~Kheffache, and E.~Osipov, ``Integer echo state
  networks: Efficient reservoir computing for digital hardware,'' 2020.
  [Online]. Available: \url{https://arxiv.org/abs/1706.00280}
\BIBentrySTDinterwordspacing

\bibitem{bitserialnn}
\BIBentryALTinterwordspacing
A.~F. Murray, A.~V.~W. Smith, and Z.~F. Butler, ``Bit-serial neural networks,''
  in \emph{Neural Information Processing Systems}, D.~Z. Anderson, Ed.\hskip
  1em plus 0.5em minus 0.4em\relax American Institute of Physics, 1988, pp.
  573--583. [Online]. Available:
  \url{http://papers.nips.cc/paper/27-bit-serial-neural-networks.pdf}
\BIBentrySTDinterwordspacing

\bibitem{cusparse}
NVIDIA, \emph{NVIDIA cuSparse API Reference}, 2020 (accessed Sept 4, 2020),
  \url{https://docs.nvidia.com/cuda/cusparse/index.html}.

\bibitem{res_pca}
\BIBentryALTinterwordspacing
B.~Penkovsky, L.~Larger, and D.~Brunner, ``Efficient design of hardware-enabled
  reservoir computing in fpgas,'' \emph{CoRR}, vol. abs/1805.03033, 2018.
  [Online]. Available: \url{http://arxiv.org/abs/1805.03033}
\BIBentrySTDinterwordspacing

\bibitem{sigma}
E.~{Qin}, A.~{Samajdar}, H.~{Kwon}, V.~{Nadella}, S.~{Srinivasan}, D.~{Das},
  B.~{Kaul}, and T.~{Krishna}, ``Sigma: A sparse and irregular gemm accelerator
  with flexible interconnects for dnn training,'' in \emph{2020 IEEE
  International Symposium on High Performance Computer Architecture (HPCA)},
  2020, pp. 58--70.

\bibitem{reddi2020mlperf}
\BIBentryALTinterwordspacing
V.~J. Reddi, C.~Cheng, D.~Kanter, P.~Mattson, G.~Schmuelling, C.~Wu,
  B.~Anderson, M.~Breughe, M.~Charlebois, W.~Chou, R.~Chukka, C.~Coleman,
  S.~Davis, P.~Deng, G.~Diamos, J.~Duke, D.~Fick, J.~S. Gardner, I.~Hubara,
  S.~Idgunji, T.~B. Jablin, J.~Jiao, T.~S. John, P.~Kanwar, D.~Lee, J.~Liao,
  A.~Lokhmotov, F.~Massa, P.~Meng, P.~Micikevicius, C.~Osborne, G.~Pekhimenko,
  A.~T.~R. Rajan, D.~Sequeira, A.~Sirasao, F.~Sun, H.~Tang, M.~Thomson, F.~Wei,
  E.~Wu, L.~Xu, K.~Yamada, B.~Yu, G.~Yuan, A.~Zhong, P.~Zhang, and Y.~Zhou,
  ``Mlperf inference benchmark,'' \emph{CoRR}, vol. abs/1911.02549, 2019.
  [Online]. Available: \url{http://arxiv.org/abs/1911.02549}
\BIBentrySTDinterwordspacing

\bibitem{laconic}
\BIBentryALTinterwordspacing
S.~Sharify, A.~D. Lascorz, M.~Mahmoud, M.~Nikolic, K.~Siu, D.~M. Stuart,
  Z.~Poulos, and A.~Moshovos, ``Laconic deep learning inference acceleration,''
  in \emph{Proceedings of the 46th International Symposium on Computer
  Architecture}, ser. ISCA '19.\hskip 1em plus 0.5em minus 0.4em\relax New
  York, NY, USA: Association for Computing Machinery, 2019, p. 304–317.
  [Online]. Available: \url{https://doi.org/10.1145/3307650.3322255}
\BIBentrySTDinterwordspacing

\bibitem{bit-fusion}
\BIBentryALTinterwordspacing
H.~Sharma, J.~Park, N.~Suda, L.~Lai, B.~Chau, J.~K. Kim, V.~Chandra, and
  H.~Esmaeilzadeh, ``Bit fusion: Bit-level dynamically composable architecture
  for accelerating deep neural networks,'' \emph{CoRR}, vol. abs/1712.01507,
  2017. [Online]. Available: \url{http://arxiv.org/abs/1712.01507}
\BIBentrySTDinterwordspacing

\bibitem{cyclic-res}
E.~S. Skibinsky-Gitlin, M.~L. Alomar, C.~F. Frasser, V.~Canals, E.~Isern,
  M.~Roca, and J.~L. Rossell{\'o}, ``Cyclic reservoir computing with fpga
  devices for efficient channel equalization,'' in \emph{Artificial
  Intelligence and Soft Computing}, L.~Rutkowski, R.~Scherer, M.~Korytkowski,
  W.~Pedrycz, R.~Tadeusiewicz, and J.~M. Zurada, Eds.\hskip 1em plus 0.5em
  minus 0.4em\relax Cham: Springer International Publishing, 2018, pp.
  226--234.

\bibitem{tan2020efficientnet}
\BIBentryALTinterwordspacing
M.~Tan and Q.~V. Le, ``Efficientnet: Rethinking model scaling for convolutional
  neural networks,'' 2020. [Online]. Available:
  \url{https://arxiv.org/abs/1905.11946}
\BIBentrySTDinterwordspacing

\bibitem{TANAKA2019100}
\BIBentryALTinterwordspacing
G.~Tanaka, T.~Yamane, J.~B. Héroux, R.~Nakane, N.~Kanazawa, S.~Takeda,
  H.~Numata, D.~Nakano, and A.~Hirose, ``Recent advances in physical reservoir
  computing: A review,'' \emph{Neural Networks}, vol. 115, pp. 100 -- 123,
  2019. [Online]. Available:
  \url{http://www.sciencedirect.com/science/article/pii/S0893608019300784}
\BIBentrySTDinterwordspacing

\bibitem{ultrascale-plus}
Xilinx, \emph{Virtex UltraScale+}, 2020 (accessed Sept 4, 2020),
  \url{https://www.xilinx.com/products/silicon-devices/fpga/virtex-ultrascale-plus.html}.

\end{thebibliography}

\end{document}